\documentclass[11pt,a4paper,oneside]{article}
\usepackage[utf8]{inputenc}
\usepackage[T2A,T1]{fontenc}
\usepackage{titlesec}
\usepackage{amsmath}
\usepackage{amsfonts}
\usepackage{mathtools}
\usepackage{amssymb}
\usepackage{graphicx}
\usepackage{subcaption}
\usepackage{amsthm}
\usepackage{setspace}
\usepackage{pgfplots}
\usepackage{fullpage}
\usepackage{exptex}
\usepackage{expthmi}
\usepackage{cite}
\usepackage{chicago}
\usepackage{rotating}
\usepackage{threeparttable}
\usepackage{multirow}
\usepackage{changepage}
\rmfamily
\makeindex
\makeatletter
\titleformat{\section}
{\normalfont\large\bfseries}{\thesection}{1em}{}

\begin{document}

\author{Alberto Baccini \and Lucio Barabesi \and Giuseppe De Nicolao}

\title{On the agreement between bibliometrics and peer review: evidence from the Italian research assessment exercises}
\author{Alberto Baccini \and Lucio Barabesi \and Giuseppe De Nicolao}
\author{Alberto Baccini$^1$ \footnote{Alberto Baccini is the recipient of grants by the Institute For New Economic Thinking Grant Institute For New Economic Thinking Grant ID INO17-00015 and INO19-00023.}\and Lucio Barabesi$^1$\and Giuseppe De Nicolao $^2$}
\date{%
    $^1$Department of Economics and Statistics, University of Siena, Italy\\%
    $^2$University of Pavia, Department of Electrical, Computer and Biomedical Engineering, Pavia, Italy\\[2ex]%}
    %\today
}
%\date{}
\maketitle

\section*{Abstract}
\newcommand{\pkg}[1]{{\normalfont\fontseries{b}\selectfont #1}}
\let\proglang=\textsf
\let\code=\texttt
This paper appraises the concordance between bibliometrics and peer review, by drawing evidence from the data of two experiments realized by the Italian governmental agency for research evaluation. The experiments were performed for validating the dual system of evaluation, consisting in the interchangeable use of bibliometyrics and peer review, adopted by the agency in the research assessment exercises. The two experiments were based on stratified random samples of journal articles. Each article was scored by bibliometrics and by peer review. The degree of concordance between the two evaluations is then computed. 

The correct setting of the experiments is defined by developing the design-based estimation of the Cohen's kappa coefficient and some testing procedures for assessing the homogeneity of missing proportions between strata. The results of both experiments show that for each research areas of hard sciences, engineering and life sciences, the degree of agreement between bibliometrics and peer review is -- at most -- weak at an individual article level. Thus, the outcome of the experiments does not validate the use of the dual system of evaluation in the Italian research assessments. More in general, the very weak concordance indicates that metrics should not replace peer review at the level of individual article. Hence, the use of the dual system of evaluation for reducing costs might introduce unknown biases in a research assessment exercise.

\textbf{Keywords:} bibliometrics \and peer review \and research assessment exercise \and Cohen's kappa \and Gwet's kappa \and design-based estimation \and stratified random sampling \and testing homogeneity of missing proportions.
\newpage

\section{Introduction}
\label{intro}

Efficient implementation of a research assessment exercise is a common challenge for policy makers. Even if attention is limited to scientific quality or scientific impact, there is a trade-off between the quality of information produced by a research assessment and its costs. Until now, two models have prevailed (Hicks, 2012): a first model based on peer review, such as the British Research Excellence Framework (REF), and a second model based on bibliometric indicators, such as Australian Excellence in Research (ERA), for the years preceding 2018. The first model is considered more costly than the second. 
In the discussion on the pros and cons of the two models, a central topic deals with the agreement between bibliometrics and peer review. Most part of the scholarly works has analyzed the REF by adopting a post-assessement perspective (Traag and Waltman, 2019). Indeed, results of the REF at various levels of aggregation are compared with those obtained by using bibliometric indicators. A clear statistical evidence on the concordance of bibliometrics and peer review would represent a very strong argument in favor of the substitution of the latter with the former. Indeed, the claim for such a substitution -- based on agreement and minor costs -- could likely appear pragmatic and hence more acceptable for academics than the argument based on juxtaposition of “objective bibliometric data” and “subjective peer reviews” (among others, see e.g. Kulczycki et al., 2017).
 
However, there are two problems hindering the adoption of the bibliometric model for research assessment. The first is how to handle the scientific fields for which bibliometrics is not easily applicable, namely social sciences and humanities. The second is how to manage research outputs covered in bibliographic databases, such as books or articles in national languages. In these cases, no substitution is possible and peer review appears as the unique possible tool for evaluating research outputs.

As a consequence, a third model of research assessment has emerged, where bibliometrics and peer review are jointly adopted: some research outputs are evaluated by bibliometrics and others by peer review. The evaluations produced by the two techniques are subsequently mixed together for computing synthetic indicators at various levels of aggregation. The Italian governmental agency for research evaluation (ANVUR) applied extensively this model in its research assessment exercises (VQR), and called it “dual system of evaluation" (Ancaiani et al., 2015). In reference to this model, the question of the agreement between bibliometrics and peer review has a constitutive nature. Indeed, a high agreement would ensure that final results of a research assessment -- at each possible level of aggregation -- are not biased by the adoption of two different instruments of evaluation. In the simplest scenario, this issue might happen when bibliometrics and peer review produce scores which substantially agree, for instance, when the research outputs evaluated by bibliometrics receive the same score by peer review -- except for random errors. In contrast, let us consider a second scenario where scores produced by bibliometrics and peer review do not agree: for instance, bibliometrics produces scores systematically lower or higher than peer review. In this more complex case, the disagreement might not be a problem solely if the two systems of evaluation are distributed homogeneously, e.g. at random, among units of assessment. Even if the concordance is not accurate at the individual article level, the errors may offset at an aggregate level (see Pride and Knoth, 2018, Traag and Waltman, 2019). In sum, the agreement between bibliometrics and peer review is functional for validating results of the assessment.

ANVUR tried to validate the use of the dual system of evaluation by implementing two extensive experiments on the agreement between bibliometrics and peer review, for each national research assessment of the years 2004-2010 (VQR1) and 2011-2014 (VQR2). The two experiments are hereinafter indicated as EXP1 and EXP2, respectively. They consisted in evaluating a random sample of articles by using both bibliometrics and peer review, and, subsequently, in assessing their degree of agreement at an individual publication level. ANVUR presented the results of EXP1 and of EXP2 as the evidence of a substantial concordance between bibliometrics and peer review. In turn, this agreement would validate the use of the dual system of evaluation and the final results of the research assessements. 

Two of the authors of the present paper documented the flaws of EXP1 and contested the interpretation of data as indicative of a substantial agreement (Baccini and De Nicolao, 2016a, 2016b, 2017a, 2017b). The present paper, conversely, takes advantage of the recent availability of the raw data of the two experiments, in order to deepen the analysis and reach conclusive results on issues that had remained open due to the sole availability of aggregated data. Therefore, this paper aims to replicate the ANVUR analysis in order to draw a solid evidence on the concordance between bibliometrics and peer review.

The paper is organized as follows. In Section 2 the literature on the two Italian experiments is framed in the general discussion on the agreement between bibliometrics and peer review. Section 3 presents the structure of EXP1 and EXP2 by reminding the essential issues of the Italian research assessment exercises. Section 4 introduces the main research questions on the sampling design and the measures of agreement. In Section 5 the correct framework for the design-based estimation of the Cohen's kappa coefficient is developed. Section 6 presents the estimates of Cohen's kappa coefficients for EXP1 and EXP2, by comparing the current results with ANVUR's findings. In Section 7, a further problem with missing data in EXP2 is presented and the homogeneity of missing proportions between scientific areas is assessed. Section 8 discusses results and concludes with some suggestions for research evaluation policy.

\section{A short review of the literature}

Most part of the literature on the agreement between bibliometrics and peer review considers the British REF. Overviews of this literature are provided by Wouters et al. (2015), Pride and Knoth (2018) and Traag and Waltman (2019). It is therefore possible to limit the discussion to a central issue which is functional to the development of this paper.
By and large, results on agreement do not converge when different approaches and statistical tools are used. Notably, the analysis conducted by the Higher Education Funding Council for England (HEFCE) in the so-called \textit{Metric Tide} report “has shown that individual metrics give significantly different outcomes from the REF peer review process, showing that metrics cannot provide a like-for-like replacement for REF peer review" (HEFCE, 2015). This analysis was performed at an individual article level by comparing the quality profile attributed by peer reviews to a set of bibliometric indicators for articles submitted to REF. 
Traag and Waltman (2019) criticized results of the \textit{Metric Tide} report by arguing that the individual publication level “is not appropriate in the context of REF". They claimed that the appropriate level is the institutional one, since “the goal of the REF is not to assess the quality of individual publications, but rather to assess ‘the quality of research in UK higher education institutions'. Therefore, the question should not be whether the evaluation of individual publications by peer review can be replaced by the evaluation of individual publications by metrics but rather whether the evaluation of institutions by peer review can be replaced by the evaluation of institutions by metrics".
In a similar vein, Pride and Knoth (2018) documented that a high concordance between bibliometric and peer-review indicators for REF is achieved when the analysis is conducted at an institutional level.

These claims should be framed in a “post-assessment" perspective, where the issue at stake is to verify the coherence between results obtained by applying one evaluative technique or the other at the desired institutional level. In the case of REF the coherence to be verified is between the adopted technique, i.e. peer review, and the alternative, i.e. bibliometrics. This viewpoint is very different from that developed in the Italian experiments and considered in this paper. In the present case, the question is whether it is possible to interchangeably use bibliometrics and peer review at an individual article level. To this end, the analysis of the agreement between bibliometrics and peer review at the level of individual publications is therefore fully justified. In turn, Traag and Waltman (2019) support the study of the concordance at an individual publication level when the issue is the possibility that bibliometrics replaces peer review at an individual level. In reference to \textit{Metric Tide} report, they explicitly wrote that “the analysis at the level of individual publications is very interesting. The low agreement at the level of individual publications supports the idea that metrics should generally not replace peer review in the evaluation of a single individual publication" (Traag and Waltman, 2019). 

As anticipated, ANVUR implemented EXP1 and EXP2 in order to justify the use of a dual system of evaluation in VQR1 and VQR2. 
As to EXP1, results were initially published as part of the official report of the research assessment exercise (ANVUR, 2013). In the official report results are synthesized by stating that “there is a more than adequate concordance between evaluation carried out through peer reviews and through bibliometrics. This results fully justifies the choice ... to use both techniques of assessment" (ANVUR, 2013).\footnote{Excerpt from Appendix B, pp. 25-26, translation by the authors. See also Baccini and De Nicolao (2016a).} Ancaiani et al. (2015) republished the complete results of EXP1, by claiming a “fundamental agreement" between bibliometrics and peer review “supporting"  the choice of using both techniques in the VQR1. Moreover, they also interpreted the experiment as indicating that “combining evaluations obtained with peer review and bibliometric methods can be considered more reliable than the usual practice of combining two or more different evaluations obtained by various reviewers of the same article".

The specific results obtained in EXP1 for the field of Economics and Statistics were largely disseminated. Bertocchi and coauthors published as far as five identical working papers where they interpreted the results of EXP1 by claiming that bibliometrics and peer review “are close substitutes” (for all, Bertocchi et al., 2013). In the version finally published in a scholarly journal, they concluded that “the agencies that run these evaluations could feel confident about using bibliometric evaluations and interpret the results as highly correlated with what they would obtain if they performed informed peer review” (Bertocchi et al., 2015). 

The results and the interpretation of EXP1 were challenged by two of the authors of the present paper on the basis of published data only, since they were unable to access raw data at the time undisclosed by ANVUR (the whole thread of papers, comments and replies includes Baccini and De Nicolao, 2016a, 2016b, 2017a, 2017b, Benedetto et al., 2017, Bertocchi et al., 2016).
The first critical appraisal was about the interpretation of the degree of concordance. Baccini and De Nicolao (2016a, 2016b) argued that, according to the available statistical guidelines, the degree of concordance between bibliometrics and peer review has to be interpreted as “unacceptable" or “poor" for all the considered research fields. The unique exception -- confirmed by a statistical meta-analysis of the data -- was Economics and Statistics, for which the protocol of the experiment was substantially modified with respect to the other fields. Baccini and De Nicolao (2017a, 2017b) also raised some questions on the sampling protocol used for EXP1, which are considered in details also in this paper. 

As for to EXP2, the results were published in the official report (ANVUR, 2017) and presented in a conference (Alfò et al., 2017). The synthesis of the results apparently confirmed the outcome of EXP1. The results of EXP2, summarized in the conclusion of the report, state that there is a “not-zero correlation”  “between peer review evaluation and bibliometric evaluation". The degree of agreement is “modest but significant. Of particular importance is the result that the degree of concordance (class and inter-row) between the bibliometric evaluation and the peer evaluation is always higher than the one existing between the two individual peer reviews" (ANVUR, 2017).\footnote{Excerpt from Appendix B, p. 33, translation by the authors.} These results are interpreted by Alfò et al. (2017) as indicating that “the combined used of bibliometric indicators for citations and journal impact may provide a useful proxy for peer review judgements” (Alfò et al., 2017). 

As anticipated, this paper aims to draw definitive evidence from the two experiments. This analysis is possible since ANVUR accepted to disclose the anonymized individual data of both EXP1 and EXP2.\footnote{The mail from Alberto Baccini to Prof. Paolo Miccoli (President of ANVUR) containing the request is dated March 12th 2019. The decision of disclosing the data was communicated by mail dated March 26th 2019. Access to the data was open on April 9th 2019.} 
It is therefore possible to replicate the results of EXP1 and EXP2, by verifying in details ANVUR methods and calculations.\footnote{Replication is solely possible at the research area levels, since -- according to a communication dated 16th March 2019 -- the data for the sub-areas “are no longer available” in the ANVUR archives.} 
For a correct understanding of the research questions, the following section presents a description of EXP1 and EXP2 in the context of the Italian research assessments.

 %For example, Research Assessment Designers (RADs) would welcome a definite proof of the agreement of peer review with some kind of bibliometric indicators -- since simpler, cheaper and more “objective” bibliometric indicators could wholly replace the peer review (Pride and Peter, 2018). 
%Many RADs and scholars would welcome that definite proof in order to justify  Again, in a final report of the research assessment in Italy, some members of the panel claim “the need of strongly reducing the peer evaluation, since it introduces a subjectivity representing a bias that cannot be normalized” (ANVUR, 2017, Area 7 report, p.113). Others RADs would welcome the definite proof of the agreement for a less radical and more practical reason: if bibliometrics and peer review agree, it is possible to combine them in a universal research assessment where some disciplines -- notably the humanities -- cannot simply be evaluated by indicators.}

\section{A brief description of the Italian experiments}

EXP1 and EXP2 were designed and performed during VQR1 and VQR2, respectively. Italian research assessement exercises aimed to evaluate research institutions, research areas and fields, both at national and institutional level (i.e. universities and departments). Synthetic indicators were obtained by aggregating the scores received by the research outputs submitted by the institutions.\footnote{Each research work was evaluated as Excellent (A, score 1), Good (B, score 0.8), Acceptable (C, score 0.5), Limited (D, score 0) in VQR1, and as Excellent (A, score 1), Elevated (B, score 0.7), Fair (C, score 0.4), Acceptable (D, score 0.1), Limited (E, score 0) in VQR2.} All the researchers with a permanent position had to submit a fixed number -- with few exceptions -- of research outputs (3 in VQR1 and 2 in VQR2). VQR1 and VQR2 were organized in 16 research area panels. Research areas were distinguished between “bibliometric areas", i.e. hard sciences, engineering and life sciences,\footnote{Mathematics and Informatics (Area 1), Physics (Area 2), Chemistry (Area 3), Earth Sciences (Area 4), Biology (Area 5), Medicine (Area 6), Agricultural and Veterinary Sciences (Area 7), Civil Engineering (Area 8b), Industrial and Information Engineering (Area 9).}
 and “non bibliometric areas", i.e. social science and humanities.\footnote{Architecture (Area 8a) Antiquities, Philology, Literary studies, Art History (Area 10), History, Philosophy, Pedagogy and Psychology (Areas 11a and 11b), Law (Area 12), Economics and Statistics (Area 13), Political and Social Sciences (Area 14).}

Both research assessments performed evaluations of the submitted research outputs by using a "dual system of evaluation", i.e. a mix of bibliometric algorithms and “informed peer review" (IPR). Panels of the “bibliometric areas” evaluated research articles mainly through bibliometrics. VQR1 and VQR2 adopted two different bibliometric algorithms. Both algorithms combined the number of citations received by an article and a journal indicator, e.g. the impact factor (for a detailed description and a critical appraisal see Abramo and D’Angelo, 2016, 2017, Franceschini and Maisano, 2017). Both algorithms were built in such a way that they generated a score if the two indicators were coherent. If the two indicators gave incoherent indications, e.g. high number of citations and low impact factor or viceversa, the article was classified as “IR” and evaluated by IPR. For evaluating books, chapters and articles in not-indexed journals, all the panels used peer review. Panels of the so-called “non-bibliometric areas” evaluated all the submitted research products exclusively by IPR. Area 13 (Economics and Statistics) was an exception, since journal articles received a score according to a journal ranking directly developed by the panel.\footnote{Journals were classified in the VQR1 as Excellent, Good, Acceptable, Limited; and as Excellent, Elevated, Fair, Acceptable, Limited in the VQR2.} In sum, only bibliometric areas plus Area 13 adopted the dual system of evaluation. 

IPR was identically organized in the two research assessments. A publication was assigned to two members of the area panel, who independently chose two anonymous reviewers. The two reviewers performed the IPR of the article by using a predefined format -- slightly different between the two research assessments and also between panels in the same assessment. Then, the referee reports were received by the two members of the area panel, who formed a so-called Consensus Group (CG) for deciding the final score of the article.
 
%ANVUR coined the expression “evaluative mix” to denote this evaluative machinery and was aware that its main problem is the possible bias induced by the adoption of different evaluation techniques. Indeed, if IPR produced scores systematically different from the ones produced by bibliometrics, this might have introduced a systematic bias in the scoring system used for ranking institutions. ANVUR precisely realized EXP1 and EXP2 for addressing that problem: a good agreement between the bibliometric evaluation and the evaluation performed by IPR might justify the adoption of the two different evaluation methods and preserve the comparability of results among areas, institutions, departments and research fields. Positive results of EXP1 and EXP2 were crucial for the soundness of Italian research assessment results: if bibliometrics and peer review do not agree and give significantly different results, the average scores of an institution might be distorted by the different percentage of scores attributed by bibliometrics and by IPR.

In order to validate the dual system of evaluation, EXP1 and EXP2 considered only the “bibliometric areas” plus Area 13. They had an identical structure. The rationale of both experiments was very simple: a sample of the journal articles submitted to the research assessment was scored by the two methods of evaluations, i.e. through the bibliometric algorithm and IPR. In such a case, IPR was implemented by involving two reviewers, according to the same rules adopted in the research assessment. These raw data were then used for analyzing (i) the agreement between the evaluation obtained through IPR and bibliometric algorithms and (ii) the agreement between the scores decided by the two reviewers. The agreement between the scores is computed by using the weighted Cohen’s kappa coefficient (Cohen, 1960), a popular index of inter-rater agreement for nominal categories (see e.g. Sheskin, 2003). A high level of agreement between IPR and bibliometric scores was interpreted as validating the dual method of evaluation.

EXP1 and EXP2 differed only for a different timing of realization. EXP1 was done simultaneously with VQR1. Hence, the reviewers were unaware of partecipating to EXP1. Indeed, they were unable to distinguish between papers of the EXP1 sample and those they had to evaluate for the research assessment. The unique exception was Area 13, where panelists and referees knew that all the journal articles belonged to the EXP1 sample -- since all the journal articles for the research assessment were evaluated automatically according to the journal ranking (Baccini and De Nicolao, 2016a). In contrast, EXP2 started after the conclusion of the activities of the research assessment. Therefore, panelists and reviewers knew that they were partecipating to EXP2. 
A second consequence of the different timing was that in EXP1 all the papers of the sample were peer-reviewed, since the successful administrative completion of the research assessment required the evaluation of all submitted articles. On the contrary, in EXP2 some papers did not receive a peer-review evaluation since some reviewers refused to do it. Therefore, in EXP2 there were missing data in the sample, which were not accounted for by ANVUR when the concordance indexes were computed.

\section{Data, sampling design and measures of agreement}

The replication of ANVUR's computations is surely useful, albeit not sufficient to reach a complete appreciation of the results of the two experiments. Indeed, some research questions should be carefully addressed. For EXP1 and EXP2, ANVUR (2013, 2017, Appendix B) adopted a stratified random sampling, where the target population was constituted by the journal articles submitted to the two research assessments. The sizes of the article populations in
EXP1 and EXP2 are $99$,005 and $77$,159, respectively. 
The sample size was about 10\% of the population size, i.e. 9,199 and 7,667 articles for EXP1 and EXP2, respectively. The stratified random samples were proportionally allocated with respect to the dimensions of the research areas. The sizes of the strata in EXP1 and EXP2 are reported in Tables 1 and 2.\footnote{Indeed, the Final Reports remark that: “The sample was stratified according to the distribution of the products among the sub-areas of the various areas” (ANVUR, 2017, Appendix B, p.1 our translation). For EXP1 results were published at a sub-area level, while for EXP2 results were solely published for areas. Moreover, the raw data at the sub-area level are not yet available.}

\begin{table}[!]
\caption{Population, sample and sub-sample sizes for scientific areas in EXP1.}
\centering
\begin{threeparttable}[b]
\scriptsize
\begin{tabular}[t]{lccc}
\hline
Scientific Areas&Population&Sample&Sub-sample\\
\hline
Area 1 - Mathematics and Informatics&$6758$&$631$&$438$\\
Area 2 - Physics&$15029$&$1412$&$1212$\\
Area 3 - Chemistry&$10127$&$927$&$778$\\
Area 4 - Earth Sciences&$5083$&$458$&$377$\\
Area 5 - Biology&$14043$&$1310$&$1058$\\
Area 6 - Medicine&$21191$&$1984$&$1602$\\
Area 7 - Agricultural and Veterinary Sciences&$6284$&$532$&$425$\\
Area 8a - Civil Engineering&$2460$&$225$&$198$\\
Area 9 - Industrial and Information Engineering&$12349$&$1130$&$919$\\
Area 13 - Economics and Statistics&$5681$&$590$&$590$\\
\hline
&$99005$&$9199$&$7597$\\
\hline
\end{tabular}
\begin{tablenotes}
\item []Source: ANVUR (2013, Appendix B). 
\end{tablenotes}
\end{threeparttable}
\end{table}

\begin{table}[!]
\caption{Population, sample, sub-sample sizes and number of missing articles for scientific areas in EXP2.}
\centering
\begin{threeparttable}[b]
\scriptsize
\scalebox{0.9}{
\begin{tabular}[t]{lcccc}
\hline
Scientific Areas&Population&Sample&Sub-sample&Missing\\
\hline
Area 1 - Mathematics and Informatics&$4631$&$467$&$344$&$23$\\
Area 2 - Physics&$10182$&$1018$&$926$&$10$\\
Area 3 - Chemistry&$6625$&$662$&$549$&$9$\\
Area 4 - Earth Sciences&$3953$&$394$&$320$&$6$\\
Area 5 - Biology&$10423$&$1037$&$792$&$86$\\
Area 6 - Medicine&$15400$&$1524$&$1071$&$231$\\
Area 7 - Agricultural and Veterinary Sciences&$6354$&$638$&$489$&$8$\\
Area 8b - Civil Engineering&$2370$&$237$&$180$&$3$\\
Area 9 - Industrial and Information Engineering&$9930$&$890$&$739$&$108$\\
Area 11b - Psychology&$1801$&$180$&$133$&$5$\\
Area 13 - Economics and Statistics&$5490$&$512$&$498$&$14$\\
\hline
&$77159$&$7667$&$6041$&$503$\\
\hline
\end{tabular}}
\begin{tablenotes}
\item []Source: ANVUR (2017, Appendix B). 
\end{tablenotes}
\end{threeparttable}
\end{table}%

A first research question deals with the statistical methodology adopted in the experiments. From this perspective, the two experiments were actually implemented in a design-based framework. Hence, their analysis requires a correct inferential setting in order to obtain the estimates of the considered concordance measures. To this aim, in Section 4 the design-based estimation of the weighted Cohen's kappa coefficient is originally developed. On the basis of these theoretical results, it is possible to check if ANVUR's estimates of agreement are correct. In particular, ANVUR's estimates of Cohen's kappa coefficients and the corresponding confidence intervals may be compared with the appropriate design-based counterparts.

ANVUR computed the final results of EXP1 and EXP2 by solely considering a sub-sample of articles - and not the whole sample. Indeed, ANVUR dropped from the computation of the concordance indexes the articles with an inconclusive bibliometric score IR, i.e. those receiving an IPR evaluation albeit they were not considered for agreement estimation.\footnote{For EXP1, the reduction of the sample due to the exclusion of the paper classified as IR was not disclosed neither in ANVUR’s official reports nor in Ancaiani et al. (2015).} Tables 1 and 2 reports the sizes of the sub-samples for EXP1 and EXP2, respectively. The exclusion of the IR papers might have boosted the value of the agreement measures, as argued by Baccini and De Nicolao (2017a). The conjecture sounds as follows. ANVUR removed from EXP1 the most problematic articles for which the bibliometric algorithm was unable to reach a score. It cannot be excluded that these articles were also particularly difficult to evaluate for peer reviewers. Hence, ANVUR calculated the agreement indicators on sub-samples of articles that were “more favorable” to agreement than the complete samples. 

The second research question, therefore, deals with the adoption of concordance measures which take into account the sizes of the IR articles which ANVUR dropped, as well as the sizes of missing articles. Actually, these articles could be ideally considered as belonging to a rating category for which agreement is not required. In such a case, there exist alternative variants of the weighted Cohen's kappa, which may suitably manage this issue. Hence, in Section 5 the design-based estimation of these variants of the weighted Cohen's kappa are also developed. In turn, in Section 6 the resulting point estimates and the corresponding confidence intervals are computed for EXP1 and EXP2, respectively.

A third and last question -- which is limited to EXP2 -- deals with the distribution of missing papers per research area, i.e. those papers for which a peer review score is unavailable. As previously remarked, Table 2 reports the number of missing papers per area. Actually, drawbacks would arise if the distribution of missing articles in the sample occurred in a non-proportional way between the strata, since in this case some research areas could be more represented than others. ANVUR (2017) claimed that this was not the case. Thus, in Section 7 a new testing procedure for the homogeneity of missing proportions between strata is developed and applied to EXP2 data.

These three questions are addressed by using the raw data of the two ANVUR experiments. The articles in each database have a unique anonymous identifier. For each article, the available variables are: (i) the research area; (ii) the bibliometric score (B); (iii) the score assigned by the first reviewer (P1); (iv) the score assigned by the second reviewer (P2); (v) the syntetic peer-review score (P). Data can be downloaded from https://doi.org/10.5281/zenodo.3727460.

\section{Design-based estimation of the Cohen's kappa coefficient}

Let us assume a
fixed population of $N$ items which are classified into $c$ categories on the
basis of two ratings. For the sake of simplicity, the $c$ categories are
labeled on the set $I=\{1,\ldots,c\}$. Hence, the $j$-th item of the
population is categorized according to the first evaluation -- say $u_j\in I$
-- and the second evaluation -- say $v_j\in I$ -- for $j=1,\ldots,N$.

A commonly-adopted measure of agreement between classifications of two raters
is given by the Cohen's kappa coefficient and its weighted generalization
(Cohen, 1960, 1968). Practitioners often adopt this index in order to assess
the inter-rater agreement for categorical ratings, while its weighted
counterpart is preferred when the categories can be considered ordinal (see
e.g. Fagerland et al., 2017, p.548, Berry et al., 2018, p.596). However, the
Cohen's kappa coefficient is also criticized for some methodological
drawbacks (for more details, see Strijbos et al., 2006, and Uebersax, 1987,
among others).

In the present design-based approach, the ``population'' weighted Cohen's
kappa coefficient is defined as
\begin{equation*}
\kappa_w=\frac{p_o-p_e}{1-p_e}\text{ ,}\tag*{(1)}
\end{equation*}
where
\begin{equation*}
p_o=\sum_{l=1}^c\sum_{m=1}^cw_{lm}p_{lm}\text{ ,
}p_e=\sum_{l=1}^c\sum_{m=1}^cw_{lm}p_{l+}p_{+m}\text{ ,}
\end{equation*}
while
\begin{equation*}
p_{lm}=\frac{1}{N}\,\sum_{j=1}^N\boldsymbol{1}_{\{l\}}(u_j)\boldsymbol{1}_{%
\{m\}}(v_j)\text{ ,
}p_{l+}=\frac{1}{N}\,\sum_{j=1}^N\boldsymbol{1}_{\{l\}}(u_j)\text{ ,
}p_{+l}=\frac{1}{N}\,\sum_{j=1}^N\boldsymbol{1}_{\{l\}}(v_j)\text{ }
\end{equation*}
and $\boldsymbol{1}_B$ is the usual indicator function of a set $B$, i.e.
$\boldsymbol{1}_B(u)=1$ if $u\in B$ and $\boldsymbol{1}_B(u)=0$, otherwise.
In practice, $p_{lm}$ is the proportion of items in the population classified
into the $l$-th category according to the first rating and into the $m$-th
category according to the second rating. Similarly, $p_{l+}$ and $p_{+l}$ are
the proportions of items categorized into the $l$-th category according to
the first rating and the second rating, respectively. In addition, the
weights $w_{lm}$, with $l,m=1,\ldots,c$, are suitably chosen in order to
consider the magnitude of disagreement (see e.g. Fagerland et al., 2017,
p.551). In particular, the (usual) unweighted Cohen's kappa coefficient is
obtained from (1) when $w_{lm}=1$ if $l=m$ and $w_{lm}=0$, otherwise.  
It is worth remarking that, for estimation purposes, the Cohen's kappa coefficient (1) can be conveniently expressed as a smooth function of population totals -- i.e. the $p_{lm}$'s, the
$p_{l+}$'s and the $p_{+l}$'s.

Let us now assume that a sampling design is adopted in order to estimate
$\kappa_w$ and let us consider a sample of fixed size $n$. Moreover, let $S$
denote the set of indexes corresponding to the sampled items -- i.e. a subset
of size $n$ of the first $N$ integers -- and let $\pi_j$ be the inclusion
probability of the first order for the $j$-th item. As an example aimed to
the subsequent application, let us assume that the population is partitioned
into $L$ strata and that $N_h$ is the size of the $h$-th stratum with
$h=1,\ldots,L$. It holds that $N=\sum_{h=1}^LN_h$. If a stratified
sampling design is considered, the sample is obtained by drawing $n_h$ items
in the $l$-th stratum by means of simple random sampling without replacement
in such a way that $n=\sum_{h=1}^Ln_h$. Therefore it turns
out that $\pi_j=n_h/N_h$ if the $j$-th item is in the $h$-th stratum (see
e.g. Thompson, 1997). When a proportional allocation is adopted, it also
holds that $n_h=nN_h/N$ -- and hence it follows that $\pi_j=n/N$.

In order to obtain the estimation of $\kappa_w$, it should be noticed that 
\begin{equation*}
\widehat{p}_{lm}=\frac{1}{N}\,\sum_{j\in
S}\frac{\boldsymbol{1}_{\{l\}}(u_j)\boldsymbol{1}_{\{m\}}(v_j)}{\pi_j}\text{
, }\widehat{p}_{l+}=\frac{1}{N}\,\sum_{j\in
S}\frac{\boldsymbol{1}_{\{l\}}(u_j)}{\pi_j}\text{ ,
}\widehat{p}_{+l}=\frac{1}{N}\,\sum_{j\in
S}\frac{\boldsymbol{1}_{\{l\}}(v_j)}{\pi_j}\text{ , }
\end{equation*}
are unbiased Horvitz-Thompson estimators of the population proportions
$p_{lm}$, $p_{l+}$ and $p_{+l}$, respectively. Thus, on the basis of the
general comments on the estimation of a function of population totals provided by Demnati and Rao (2004),
a ``plug-in'' estimator of (1) is given by
\begin{equation*}
\widehat{\kappa}_w=\frac{\widehat{p}_o-\widehat{p}_e}{1-\widehat{p}_e}\text{
,}\tag*{(2)}
\end{equation*}
where
\begin{equation*}
\widehat{p}_o=\sum_{l=1}^c\sum_{m=1}^cw_{lm}\widehat{p}_{lm}\text{ ,
}\widehat{p}_e=\sum_{l=1}^c\sum_{m=1}^cw_{lm}\widehat{p}_{l+}\widehat{p}_{+m}%
\text{ .}
\end{equation*}
Even if estimator (2) is biased, its bias is negligible since (1) is a
differentiable function of the population totals with non-null derivatives
(for more details on such a result, see e.g. Thompson, 1997, p.106).

As usual, variance estimation is mandatory in order to achieve an evaluation
of the accuracy of the estimator. Since (2) is a rather involved function of
sample totals, its variance may be conveniently estimated by the
linearization method or by the jackknife technique (see e.g. Demnati and Rao,
2004, and references therein). Alternatively, a bootstrap approach -- which is
based on a pseudo-population method -- may be suitably considered (for more
details on this topic, see e.g. Quatember, 2015).

It should be remarked that inconclusive ratings occur in EXP1 and EXP2 and -- in addition -- missing ratings are also present in EXP2. However, even if ANVUR does not explicitly states
this issue, its target seems to be the sub-population of items with two
reported ratings. Hence, some suitable variants of the Cohen's kappa
coefficient have to be considered. In order to deal with an appropriate
definition of the population parameter in this setting, the three suggestions
provided by De Raadt et al. (2019) could be adopted. For the sake of
simplicity, let us suppose that inconclusive or missing ratings are
classified into the $c$-th category. A first way to manage the issue consists
in deleting all items which are not classified by both raters and apply the
weighted Cohen's kappa coefficient to the items with two ratings (see also Strijbos and Stahl,
2007). After some straightforward algebra, this variant of the population
weighted Cohen's kappa coefficient may be written as
\begin{equation*}
\kappa_w^{(1)}=\frac{p_o^{(1)}-p_e^{(1)}}{1-p_e^{(1)}}\text{ ,}\tag*{(3)}
\end{equation*}
where
\begin{equation*}
p_o^{(1)}=\frac{\sum_{l=1}^{c-1}\sum_{m=1}^{c-1}w_{lm}p_{lm}}{%
\sum_{l=1}^{c-1}\sum_{m=1}^{c-1}p_{lm}}\text{ ,
}p_e^{(1)}=\frac{\sum_{l=1}^{c-1}%
\sum_{m=1}^{c-1}w_{lm}(p_{l+}-p_{lc})(p_{+m}-p_{cm})}{(\sum_{l=1}^{c-1}%
\sum_{m=1}^{c-1}p_{lm})^2}\text{ .}
\end{equation*}
It is worth noting that (3) could be not a satisfactory index, since it
does not take into account the size of inconclusive or missing ratings.
Similarly to (1), its variant (3) can be estimated as
\begin{equation*}
\widehat{\kappa}_w^{(1)}=\frac{\widehat{p}_o^{(1)}-\widehat{p}_e^{(1)}}{1-%
\widehat{p}_e^{(1)}}\text{ ,}\tag*{(4)}
\end{equation*}
where
\begin{equation*}
\widehat{p}_o^{(1)}=\frac{\sum_{l=1}^{c-1}\sum_{m=1}^{c-1}w_{lm}%
\widehat{p}_{lm}}{\sum_{l=1}^{c-1}\sum_{m=1}^{c-1}\widehat{p}_{lm}}\text{ ,
}\widehat{p}_e^{(1)}=\frac{\sum_{l=1}^{c-1}\sum_{m=1}^{c-1}w_{lm}(%
\widehat{p}_{l+}-\widehat{p}_{lc})(\widehat{p}_{+m}-\widehat{p}_{cm})}{(%
\sum_{l=1}^{c-1}\sum_{m=1}^{c-1}\widehat{p}_{lm})^2}\text{ .}
\end{equation*}

The second proposal by De Raadt et al. (2019) for a variant of the weighted Cohen's kappa coefficient
is based on Gwet's kappa (Gwet, 2014). The population weighted Gwet's kappa
may be defined as
\begin{equation*}
\kappa_w^{(2)}=\frac{p_o^{(2)}-p_e^{(2)}}{1-p_e^{(2)}}\text{ ,}\tag*{(5)}
\end{equation*}
where
\begin{equation*}
p_o^{(2)}=\frac{\sum_{l=1}^{c-1}\sum_{m=1}^{c-1}w_{lm}p_{lm}}{%
\sum_{l=1}^{c-1}\sum_{m=1}^{c-1}p_{lm}}\text{ ,
}p_e^{(2)}=\frac{\sum_{l=1}^{c-1}%
\sum_{m=1}^{c-1}w_{lm}p_{l+}p_{+m}}{(1-p_{c+})(1-p_{+c})}\text{ .}
\end{equation*}
This index considers the sizes of inconclusive or
missing ratings. Indeed, even if $p_o^{(2)}=p_o^{(1)}$, the quantity
$p_e^{(2)}$ is actually a weighted sum of the products of type $p_{l+}p_{+l}$
-- in contrast to the quantity $p_e^{(1)}$ which is a weighted sum of the
products of type $(p_{l+}-p_{lc})(p_{+m}-p_{cm})$. In turn, (5) may be
estimated by means of
\begin{equation*}
\widehat{\kappa}_w^{(2)}=\frac{\widehat{p}_o^{(2)}-\widehat{p}_e^{(2)}}{1-%
\widehat{p}_e^{(2)}}\text{ ,}\tag*{(6)}
\end{equation*}
where
\begin{equation*}
\widehat{p}_o^{(2)}=\frac{\sum_{l=1}^{c-1}\sum_{m=1}^{c-1}w_{lm}%
\widehat{p}_{lm}}{\sum_{l=1}^{c-1}\sum_{m=1}^{c-1}\widehat{p}_{lm}}\text{ ,
}\widehat{p}_e^{(2)}=\frac{\sum_{l=1}^{c-1}\sum_{m=1}^{c-1}w_{lm}%
\widehat{p}_{l+}\widehat{p}_{+m}}{(1-\widehat{p}_{c+})(1-\widehat{p}_{+c})}%
\text{ .}
\end{equation*}

The third proposal for a variant of (1) stems on assuming null weights for
the inconclusive or missing ratings (De Raadt et al., 2019), i.e. by assuming that $w_{lm}=0$ if
$l=c$ or $m=c$. Hence, this variant is defined as
\begin{equation*}
\kappa_w^{(3)}=\frac{p_o^{(3)}-p_e^{(3)}}{1-p_e^{(3)}}\text{ ,}\tag*{(7)}
\end{equation*}
where
\begin{equation*}
p_o^{(3)}=\sum_{l=1}^{c-1}\sum_{m=1}^{c-1}w_{lm}p_{lm}\text{ ,
}p_e^{(3)}=\sum_{l=1}^{c-1}\sum_{m=1}^{c-1}w_{lm}p_{l+}p_{+m}\text{ .}
\end{equation*}
In turn, (7) may be estimated by means of
\begin{equation*}
\widehat{\kappa}_w^{(3)}=\frac{\widehat{p}_o^{(3)}-\widehat{p}_e^{(3)}}{1-%
\widehat{p}_e^{(3)}}\text{ ,}\tag*{(8)}
\end{equation*}
where
\begin{equation*}
\widehat{p}_o^{(3)}=\sum_{l=1}^{c-1}\sum_{m=1}^{c-1}w_{lm}\widehat{p}_{lm}%
\text{ ,
}\widehat{p}_e^{(3)}=\sum_{l=1}^{c-1}\sum_{m=1}^{c-1}w_{lm}\widehat{p}_{l+}%
\widehat{p}_{+m}\text{ .}
\end{equation*}

The previous findings are applied to the data collected in EXP1 and EXP2 in the following section.

\section{Cohen's kappa coefficient estimation in the Italian experiment}

The theoretical results presented in Section 5 can be applied to the raw data of the two experiments. Therefore, it is possible to implement appropriate estimates of the considered weighted Cohen's kappa coefficients for the agreement (i) between bibliometric and peer-review ratings and (ii) between the ratings of the first referee (P1) and the second referee (P2). The dot-plot graphics of the distributions of the ratings are provided as Supplementary materials, available at https://doi.org/10.5281/zenodo.3727460.

Some preliminary considerations are required on the choice of the weights for the computation of Cohen's kappa. Let ${\rm{\bf W}}=(w_{lm})$ generally denote the square matrix of order $c$ of the weights. The selection of the weights is completely subjective and the adoption of different sets of weights may obviously modify the concordance level. ANVUR presented results for two sets of weights in EXP1 and EXP2. The first set of weights consisted in the usual linear weights, i.e. $w_{lm}=1-|l-m|/(c-1)$. In such a case, the matrices of linear weights for EXP1 and EXP2 are given, respectively, by
\begin{equation*}
{\rm{\bf W}}=
\begin{pmatrix} 
1&	0.67&	0.33&	0\\
0.67&	1&	0.67&	0.33\\
0.33&	0.67&	1&	0.67\\
0&	0.33&	0.67&	1
\end{pmatrix}
\end{equation*}
and
\begin{equation*}
{\rm{\bf W}}=
\begin{pmatrix} 
1&	0.75&	0.50&	0.25&	0\\
0.75&	1&	0.75&	0.50&	0.25\\
0.50&	0.75&	1&	0.75&	0.50\\
0.25&	0.50&	0.75&	1&	0.75\\
0&	0.25&	0.50&	0.75&	1
\end{pmatrix}.
\end{equation*}
The second set was originally developed by ANVUR and named “VQR-weights". The matrices of VQR-weights for EXP1 and EXP2 are respectively given by
\begin{equation*}
{\rm{\bf W}}=
\begin{pmatrix} 
1&	0.8&	0.5&	0\\
0.8&	1&	0.7&	0.2\\
0.5&	0.7&	1&	0.5\\
0&	0.2&	0.5&	1
\end{pmatrix}
\end{equation*}
and
\begin{equation*}
{\rm{\bf W}}=
\begin{pmatrix} 
1&	0.7&	0.4&	0.1&	0\\
0.7&	1&	0.7&	0.4&	0.3\\
0.4&	0.7&	1&	0.7&	0.6\\
0.1&	0.4&	0.7&	1&	0.9\\
0&	0.3&	0.6&	0.9&	1
\end{pmatrix}.
\end{equation*}
The VQR-weights were based on the scores adopted in the research assessments even if they appear counter-intuitive, since they attribute different weights to a same category distance. For example, in EXP1 a distance of two categories is weighted with 0.5 if it occurs for the first and the third category, while it is solely weighted with 0.2 if it occurs for the  second and fourth category.
In order to reproduce ANVUR's results, the sets of linear weights and VQR-weights are solely considered. In addition, for improving readability, the analysis and the comments are limited to the computation based on VQR-weights, while the results for linear weights are available as Supplementary materials from https://doi.org/10.5281/zenodo.3727460. 

At first, the estimation of (3), (5) and (7) are considered for the agreement of the bibliometric and peer-review ratings by means of the estimators (4), (6) and (8). The estimation was carried out for each
area and for the global population in both EXP1 and EXP2. Variance estimation was carried out by means of the Horvitz-Thompson based bootstrap -- stemming on the use of a pseudo-population -- which is described by Quatember (2015, p.16, p.80). The whole computation was implemented by means of the algebraic software Mathematica (Wolfram Research Inc., 2014). The corresponding Mathematica notebooks are available on request. The point and interval estimates are given in Tables 3 and 4. The columns labeled “ANVUR" report the point and interval estimates provided by ANVUR (2013, 2017). Moreover, in Figures 1 and 2 the estimates (4), (6) and (8) and the corresponding confidence intervals at the $95\%$ confidence level are plotted in the “error-bar'' style.

Actually, the point estimates given by ANVUR correspond to those computed by means of (4). Thus, even if this issue is not explicitly stated in its reports (ANVUR, 2013, 2017), ANVUR focused on the sub-population of articles with two reported ratings and considered the estimation of (3). Hence, the Cohen's kappa coefficient assumed by ANVUR does not account for the size of inconclusive ratings in EXP1, and for the size of inconclusive or missing ratings in EXP2. Moreover, the confidence intervals provided by ANVUR -- and reported in Tables 3 and 4 -- are the same computed by means of the packages \pkg{psych} (in the case of EXP1) and \pkg{vcd} (in the case of EXP2) of the software \proglang{R} (R Core Team, 2019). Unfortunately, these confidence intervals rely on the model-based approximation for large samples described by Fleiss et al. (2003, p.610). Thus, even if ANVUR has apparently adopted a design-based inference, the variance estimation is carried out in a model-based approach. The columns corresponding to $\widehat{\kappa}_w^{(1)}$ of Tables 3 and 4 show the appropriate version of ANVUR estimates, i.e. the design-based point estimates and the corresponding confidence intervals, which were computed by the bootstrap method. These confidence intervals are generally narrower than those originally computed by ANVUR -- consistently with the fact that a stratified sampling design is carried out, rather than a simple random sampling design. 

It is also convenient to consider the two alternative definitions of the weighted Cohen's kappa coefficient (5) and (7) and the corresponding estimators (6) and (8). These concordance measures take into account the sizes of the discarded articles -- as formally explained in Section 5. From Tables 3 and 4, for both EXP1 and EXP2, the point and interval estimates corresponding to $\widehat{\kappa}_w^{(2)}$ are similar to those corresponding to $\widehat{\kappa}_w^{(1)}$. In contrast, the point and interval estimates corresponding to $\widehat{\kappa}_w^{(3)}$ tend to be sistematically smaller than those corresponding to $\widehat{\kappa}_w^{(1)}$. Arguably, this outcome should be expected. Indeed, (7) is likely to be more conservative than (3) and (5), since it assigns null weights to IR and missing articles. 
 
By considering Figure 1, the first evidence is that Area 13 -- i.e. Economics and Statistics -- is likely to be an outlier. In particular, point and interval estimates are identical when estimated by using (4), (6) or (8), since in Area 13 the use of simple journal ranking -- as remarked in Section 3 -- did not produce IR score. More importantly, in Area 13 the value of agreement for EXP1 is higher than 0.50 and much higher than the values of all the other areas. Baccini and De Nicolao (2016a, 2016b) documented that in Area 13 the protocol of the experiment was substantially modified with respect to the other areas and contributed to boost the concordance between bibliometrics and peer review. In contrast, from Figure 2, Area 13 cannot be considered an outlier as in EXP1 -- even if shows slightly higher values of agreement with respect to the other areas. Indeed, in EXP2 Area 13 adopted the same protocol of the other areas. Thus, it could be conjectured that the higher agreement was due to the exclusive use of journal ranking for attributing bibliometric scores.

Let us focus on the other areas in EXP1 and EXP2. The confidence intervals corresponding to $\widehat{\kappa}_w^{(1)}$ and $\widehat{\kappa}_w^{(2)}$ are largely overlapped. For most of the areas, the upper bound of the confidence intervals corresponding to $\widehat{\kappa}_w^{(3)}$ is smaller than the lower bound of the confidence intervals corresponding to $\widehat{\kappa}_w^{(1)}$ and $\widehat{\kappa}_w^{(2)}$. Therefore, ANVUR's choice of discarding IR and missing articles presumably boosted the agreement between bibliometrics and peer review. Anyway, the upper bounds of the confidence intervals corresponding to $\widehat{\kappa}_w^{(2)}$ are generally smaller than $0.40$, and those corresponding to $\widehat{\kappa}_w^{(3)}$ are generally smaller than $0.30$. These values indicate -- at most -- a weak value of concordance when the simple Cohen's kappa is considered -- see e.g. the recent guideline for interpreting Cohen's kappa coefficient by Fagerland et al. (2017, p.549) and the survey provided by Baccini and De Nicolao (2016a). Hence, such small values of the weighted Cohen's kappa coefficients indicate an even worse concordance.

\begin{table}[!]
\centering
	\begin{threeparttable}[b]
	\scriptsize
	\caption{Cohen's kappa coefficient estimates (percent) for EXP1 (95\%
confidence level intervals in parenthesis), bibliometric vs peer review ratings.}
	%\lcrline{}{%
	\begin{tabular}[t]{ccccc}
\hline
Area&ANVUR\tnote{a}&$\widehat{\kappa}_w^{(1)}$&$\widehat{\kappa}_w^{(2)}$&$\widehat{%
\kappa}_w^{(3)}$\\
\hline
1&31.73(23.00,40.00)&31.73(25.21,38.26)&33.4$0$(26.8$0$,40.$00$)&15.07(11.76,%
18.38)\\
2&25.15(21.00,29.00)&25.15(21.1$0$,29.19)&29.15(25.29,33.01)&18.91(16.24,%
21.58)\\
3&22.96(17.00,29.00)&22.96(18.05,27.86)&23.98(19.09,28.88)&14.52(11.32,17.71)%
\\
4&29.85(21.00,39.00)&29.85(23.32,36.37)&30.24(23.69,36.79)&20.32(15.66,24.99)%
\\
5&34.53(29.00,40.00)&34.53(30.51,38.54)&36.62(32.72,40.51)&23.85(21.13,26.58)%
\\
6&33.51(29.00,38.00)&33.51(30.30,36.72)&34.62(31.47,37.77)&22.73(20.51,24.95)%
\\
7&34.37(27.00,42.00)&34.37(27.99,40.75)&36.62(30.59,42.65)&22.60(18.43,26.77)%
\\
8a&22.61(11.00,34.00)&22.61(12.70,32.52)&22.99(13.06,32.92)&16.35(8.90,23.80)%
\\
9&17.10(13.00,21.00)&17.10(13.17,21.03)&21.95(17.78,26.11)&12.56(10.12,15.01)%
\\
13&61.04(53.00,69.00)\tnote{b}&54.17(49.37,58.98)&54.17(49.37,58.98)&54.17(49.37,%
58.98)\\
\hline
All&38.00(36.00,40.00)\tnote{c}&34.15(32.64,35.66)&35.76(34.28,37.24)&23.28(22.23,%
24.33)\\
\hline
\end{tabular}
\begin{tablenotes}
\item [a] Source: ANVUR (2013, Appendix B). Reproduced in Ancaiani et al. (2015).
\item [b] Estimated with the wrong system of weights as documented in Baccini and De Nicolao (2017a). Benedetto et al. (2017) justified it as “factual error in editing of the table” and published a corrected estimate of 54.17.
\item [c] Ancaiani et al. (2015) reported a different estimate of 34.41, confirmed also in Benedetto et al. (2017).
\end{tablenotes}
\end{threeparttable}
\end{table}
%}

\begin{table}[!]
\centering
\begin{threeparttable}[b]
\scriptsize
\caption{Cohen's kappa coefficient estimates (percent) for EXP2 (95\%
confidence level intervals in parenthesis), bibliometric vs peer review ratings.}
%\lcrline{}{%
\begin{tabular}[t]{ccccc}
\hline
Area&ANVUR\tnote{a}&$\widehat{\kappa}_w^{(1)}$&$\widehat{\kappa}_w^{(2)}$&$\widehat{%
\kappa}_w^{(3)}$\\
\hline
1&21.50(15.10,27.80)&21.48(15.38,27.58)&22.85(16.71,29.00)&14.97(11.79,18.16)%
\\
2&26.50(22.40,30.50)&26.48(22.61,30.34)&28.66(24.86,32.46)&22.35(19.46,25.23)%
\\
3&19.50(14.30,24.70)&19.49(14.60,24.38)&20.85(16.01,25.69)&13.71(10.71,16.72)%
\\
4&23.90(16.60,31.20)&23.90(17.02,30.77)&24.52(17.75,31.28)&15.78(11.55,20.01)%
\\
5&24.10(19.70,28.40)&24.07(19.98,28.15)&25.01(20.97,29.05)&19.93(17.75,22.11)%
\\
6&22.80(19.50,26.20)&22.83(19.62,26.04)&24.47(21.32,27.62)&21.00(19.49,22.51)%
\\
7&27.00(21.30,32.70)&27.01(21.66,32.36)&28.76(23.56,33.96)&16.02(13.05,18.99)%
\\
8b&17.20(8.80,25.60)&17.21(9.183,25.23)&20.36(12.55,28.16)&11.45(7.23,15.67)\\
9&16.90(12.90,21.00)&16.91(13.04,20.78)&19.62(15.82,23.42)&18.51(16.58,20.44)%
\\
11b&24.10(13.70,34.50)&24.09(14.30,33.88)&25.45(15.93,34.97)&14.76(9.556,%
19.95)\\
13&30.90(26.20,35.50)&30.85(26.36,35.34)&30.85(26.36,35.34)&31.54(27.51,%
35.57)\\
\hline
All&26.00(24.50,27.60)&26.10(24.64,27.56)&27.31(25.87,28.74)&20.88(20.05,%
21.71)\\
\hline
\end{tabular}
\begin{tablenotes}
\item [a] Source: ANVUR (2017, Appendix B). 
\end{tablenotes}
\end{threeparttable}
\end{table}%
%}

\begin{figure}
 \centering	
 \includegraphics [scale=0.30]{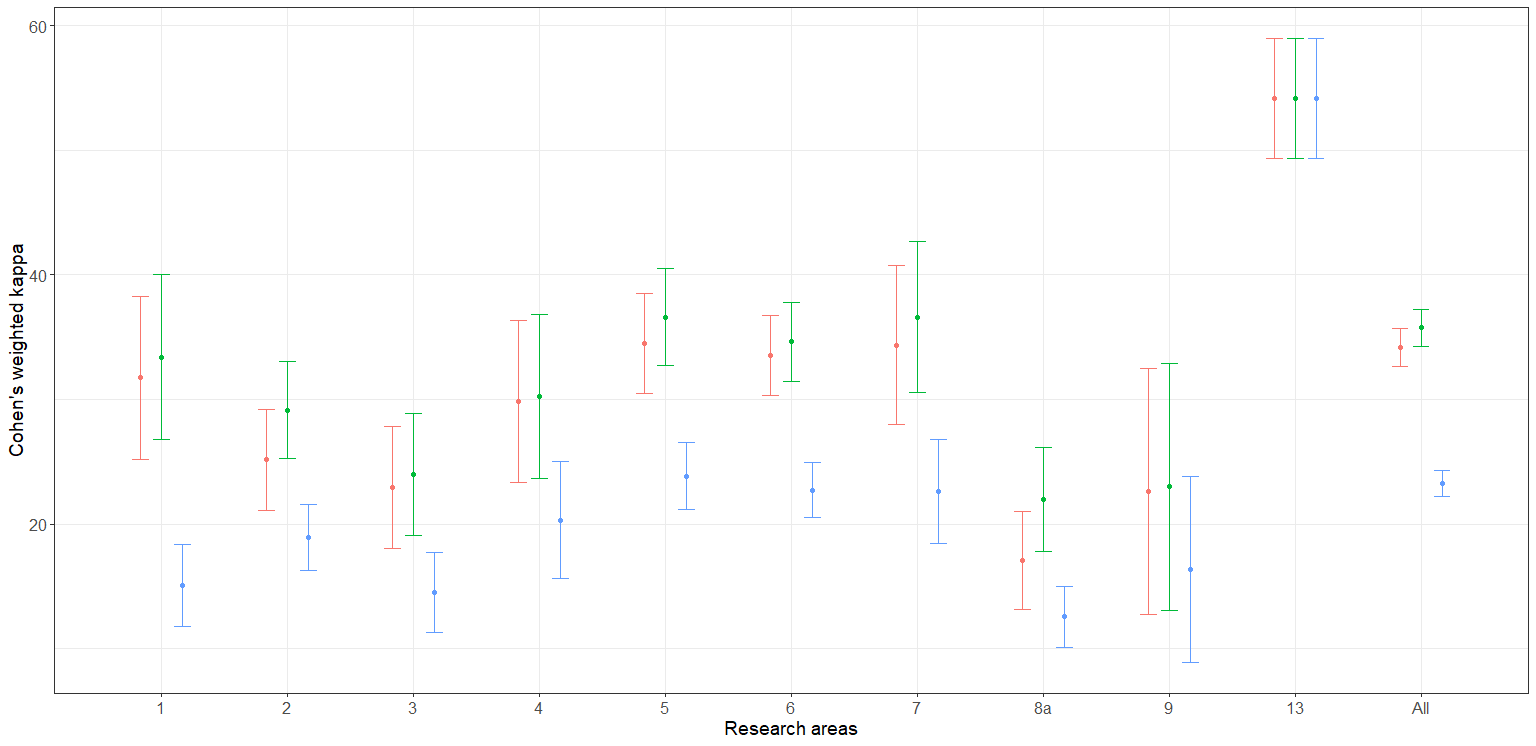} 
 \caption{“Error-bar'' plots of the Cohen's kappa coefficient estimates (percent) for EXP1, bibliometric vs peer review ratings. The confidence intervals are at $95\%$ confidence level and estimates corresponding to $\widehat{\kappa}_w^{(1)}$, $\widehat{\kappa}_w^{(2)}$ and $\widehat{\kappa}_w^{(3)}$ are in red, green and blue, respectively.}
   \label{fig:IE_ECON}
\end{figure}

\begin{figure}
 \centering	
     \includegraphics [scale=0.30]{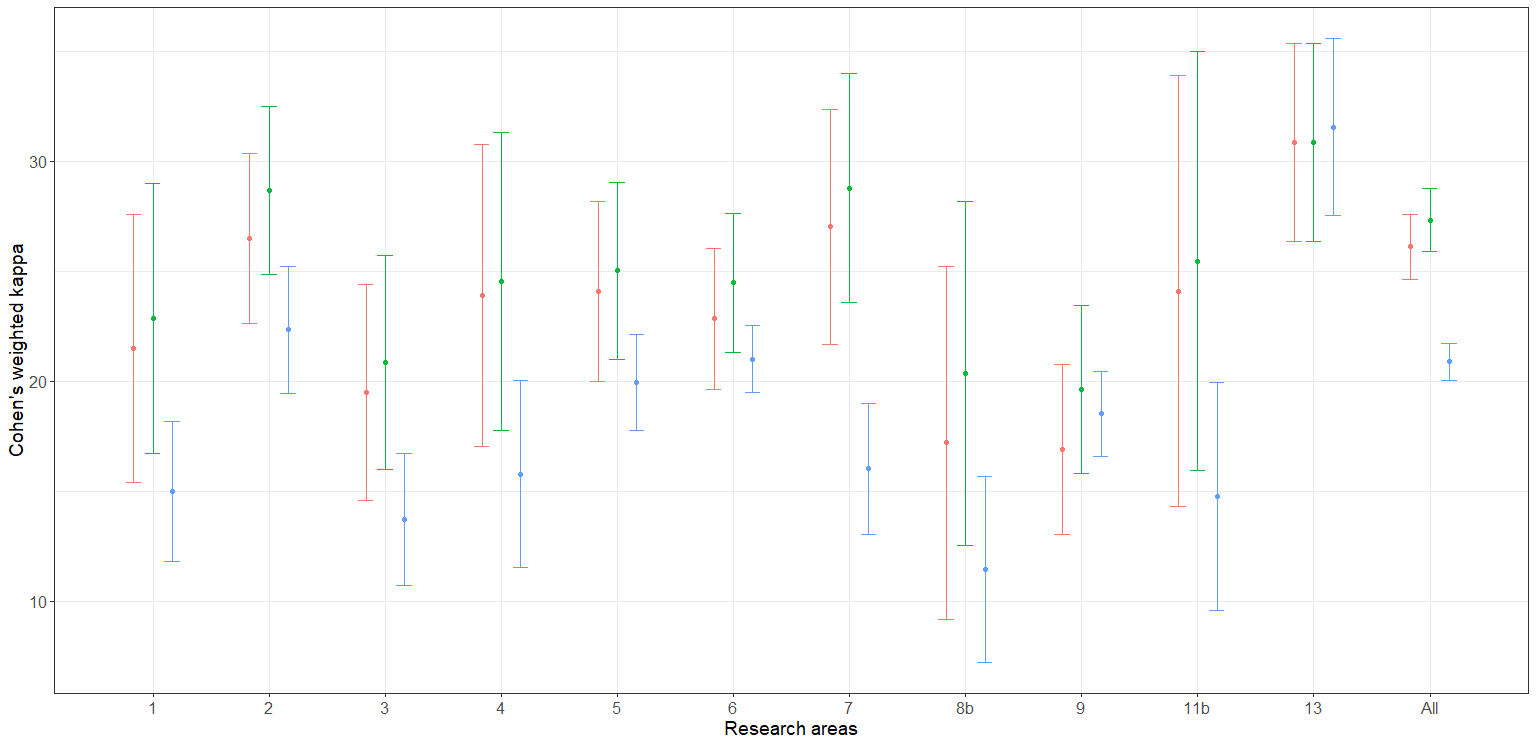} 
   \caption{“Error-bar'' plots of the Cohen's kappa coefficient estimates (percent) for EXP2, bibliometric vs peer review ratings. The confidence intervals are at $95\%$ confidence level and estimates corresponding to $\widehat{\kappa}_w^{(1)}$, $\widehat{\kappa}_w^{(2)}$ and $\widehat{\kappa}_w^{(3)}$ are in red, green and blue, respectively.}
   \label{fig:IE_ECON}
\end{figure}

Subsequently, it is also considered the estimation of the Cohen's kappa coefficient for the agreement of the ratings attributed to the articles by the two reviewers, i.e. P1 and P2.
%The point estimates and the corresponding confidence intervals are in Tables 5 and 6 for EXP1 and EXP2 respectively. The columns labeled "ANVUR" report the estimates provided by ANVUR (2013, 2017).} 
%\textcolor{red}{Figure 3 and Figure 4 draw the estimates and the corresponding confidence as error bars.} 
Thus, the estimation of (3) is computed for the population of articles, for the sub-population of articles receiving a definite bibliometric rating (DBR) and for the sub-population of articles with an inconclusive bibliometric rating (IR). The point and interval estimates are reported in Tables 5 and 6, and displayed in Figures 3 and 4 in “error-bar'' style. It should be remarked that -- owing to the use of journal ranking -- there are no IR articles for Area 13.

In Tables 5 and 6, the column labeled “ANVUR" reports the estimates provided by ANVUR (2013, 2017). In turn, ANVUR did not explicitly state that it aimed to estimate (3) in the sub-population of articles with a definite bibliometric rating. However, this issue can be inferred from Tables 5 and 6, where -- unless specific errors in the ANVUR computation for some areas -- the ANVUR point estimates correspond to $\widehat{\kappa}_w^{(1)}$ for the sub-population DBR. The confidence intervals provided by ANVUR are the same computed by means of the packages \pkg{psych} and \pkg{vcd} of the software \proglang{R} (R Core Team, 2019). Thus, in this case also, ANVUR has apparently adopted a design-based inference, even if variance estimation is carried out in a model-based approach. Therefore, in Tables 5 and 6 the column corresponding to $\widehat{\kappa}_w^{(1)}$ for the sub-population DBR reports the appropriate version of ANVUR point and interval estimates.
The point estimate of (3) between the two reviewers for the population of articles, i.e. the column corresponding to $\widehat{\kappa}_w^{(1)}$ in Tables 5 and 6, is generally lower then 0.30 with the exception of Area 13. The confidence intervals corresponding to $\widehat{\kappa}_w^{(1)}$ overlap with the confidence intervals corresponding to $\widehat{\kappa}_w^{(1)}$ for the sub-population DBR. From Figures 3 and 4, it is also apparent that $\widehat{\kappa}_w^{(1)}$ is generally greater than $\widehat{\kappa}_w^{(1)}$ for the sub-population IR. This last issue confirms the conjecture by Baccini and De Nicolao (2017a) that articles for which bibliometric rating was inconclusive were also the more difficult to evaluate for reviewers, by showing a smaller degree of agreement for these papers.

\begin{table}[!]
\centering
\begin{threeparttable}[b]
\scriptsize
\caption{Cohen's kappa coefficient estimates (percent) for EXP1 (95\%
confidence level intervals in parenthesis), P1 vs P2 ratings.}
%\lcrline{}{%
\begin{tabular}[t]{ccccc}
\hline
Area&ANVUR\tnote{a}&$\widehat{\kappa}_w^{(1)}$&$\widehat{\kappa}_w^{(1)}$ (DBR)\tnote{e}&$\widehat{\kappa}_w^{(1)}$ (IR)\tnote{f}\\
\hline
1&35.16(26.00,44.00)&33.31(27.54,39.09)&35.16(25.26,45.06)&28.87(18.17,39.57)%
\\
2&22.71(18.00,28.00)&23.42(19.44,27.41)&22.71(17.28,28.14)&19.31(9.227,29.39)%
\\
3&23.81(17.00,30.00)&20.83(16.00,25.65)&23.81(17.73,29.89)&2.56(-7.01,12.15)\\
4&25.48(15.00,36.00)&23.27(16.55,30.00)&25.48(16.59,34.36)&12.37(-3.47,28.23)%
\\
5&27.17(21.00,33.00)&24.85(20.76,28.93)&27.17(21.56,32.78)&11.12(1.77,20.46)\\
6&23.56(19.00,29.00)&21.85(18.57,25.12)&23.56(19.09,28.03)&11.84(4.19,19.48)\\
7&26.56(21.00,33.00)\tnote{b}&17.47(11.34,23.61)&16.99(8.15,25.83)&16.41(2.91,29.90)\\
8a&19.43(6.00,32.00)&19.92(9.64,30.21)&19.43(6.65,32.20)&23.77(-7.45,54.99)\\
9&18.18(12.00,24.00)&19.39(14.93,23.84)&18.18(11.72,24.64)&21.1(10.70,31.50)\\
13&45.99(38.00,54.00)\tnote{c}&38.98(33.50,44.47)&38.98(33.50,44.47)&-\\
\hline
All&33.00(31.00,35.00)\tnote{d}&26.68(25.16,28.20)&27.92(25.90,29.95)&18.90(15.30,%
22.50)\\
\hline
\end{tabular}
\begin{tablenotes}
\item [a] Source: ANVUR (2013, Appendix B). Reproduced in Ancaiani et al. (2015).
\item [b] Estimated with the wrong system of weights reported in Baccini and De Nicolao (2017a). Benedetto et al. (2017) justified it as “factual error in editing of the table” and published a corrected estimate of 16.99.
\item [c] Estimated with the wrong system of weights reported in Baccini and De Nicolao (2017a). Benedetto et al. (2017) justified it as “factual error in editing of the table” and published a corrected estimate of 38.998.
\item [d] Ancaiani et al. (2015) reported a different estimate of 28.16, confirmed also in Benedetto et al. (2017).
\item [e] Weighted Cohen's kappa for the sets of articles with a definite bibliometric rating (DBR). 
\item [f] Weighted Cohen's kappa for the sets of articles without a definite bibliometric rating and submitted to informed peer review (IPR).
\end{tablenotes}
\end{threeparttable}
\end{table}%
%}{}

\begin{table}[!]
\centering
\begin{threeparttable}[b]
\scriptsize
\caption{Cohen's kappa coefficient estimates (percent) for EXP2 (95\%
confidence level intervals in parenthesis), P1 vs P2 rating.}
%\lcrline{}{%
\begin{tabular}[t]{ccccc}
\hline
Area&ANVUR\tnote{a}&$\widehat{\kappa}_w^{(1)}$&$\widehat{\kappa}_w^{(1)}$ (DBR)\tnote{b}&$\widehat{\kappa}_w^{(1)}$ (IR)\tnote{c}\\
\hline
1&20.20(12.90,27.50)&23.92(17.68,30.15)&20.18(11.11,29.25)&35.71(22.11,49.31)%
\\
2&19.50(14.60,24.40)&21.13(16.75,25.52)&19.50(14.24,24.77)&20.29(6.81,33.77)\\
3&14.00(7.90,20.10)&14.67(9.36,19.98)&13.99(7.14,20.83)&15.53(3.04,28.02)\\
4&18.90(11.10,26.80)&18.63(11.97,25.29)&18.94(10.14,27.75)&12.4(-2.21,27.02)\\
5&19.50(14.60,24.50)&20.21(15.80,24.63)&19.53(13.65,25.41)&20.73(9.82,31.63)\\
6&19.10(17.90,23.20)&17.84(14.24,21.44)&19.08(14.29,23.87)&7.69(-0.73,16.13)\\
7&19.60(13.40,25.80)&22.38(17.14,27.63)&19.57(11.58,27.57)&28.34(17.54,39.15)%
\\
8b&3.50(-0.06,13.20)&8.70(0.22,17.19)&3.47(-9.42,16.37)&22.41(5.90,38.92)\\
9&15.10(9.90,20.30)&15.36(10.84,19.89)&15.09(8.87,21.31)&12.71(1.65,23.76)\\
11b&25.70(13.30,38.20)&25.79(15.39,36.19)&25.72(8.93,42.50)&20.68(0.22,41.15)%
\\
13&31.20(25.40,36.90)&31.15(25.69,36.61)&31.15(25.69,36.61)&-\\
\hline
All&23.40(21.60,25.20)&23.54(21.97,25.10)&23.50(21.48,25.52)&19.85(15.94,%
23.77)\\
\hline
\end{tabular}
\begin{tablenotes}
\item [a] Source: ANVUR (2017). 
\item [b] Weighted Cohen's kappa for the sets of articles with a definite bibliometric rating (DBR). 
\item [c] Weighted Cohen's kappa for the sets of articles without a definite bibliometric rating and submitted to informed peer review (IPR).
\end{tablenotes}
\end{threeparttable}
\end{table}%
%}

\begin{figure}
\centering	
\includegraphics [scale=0.30]{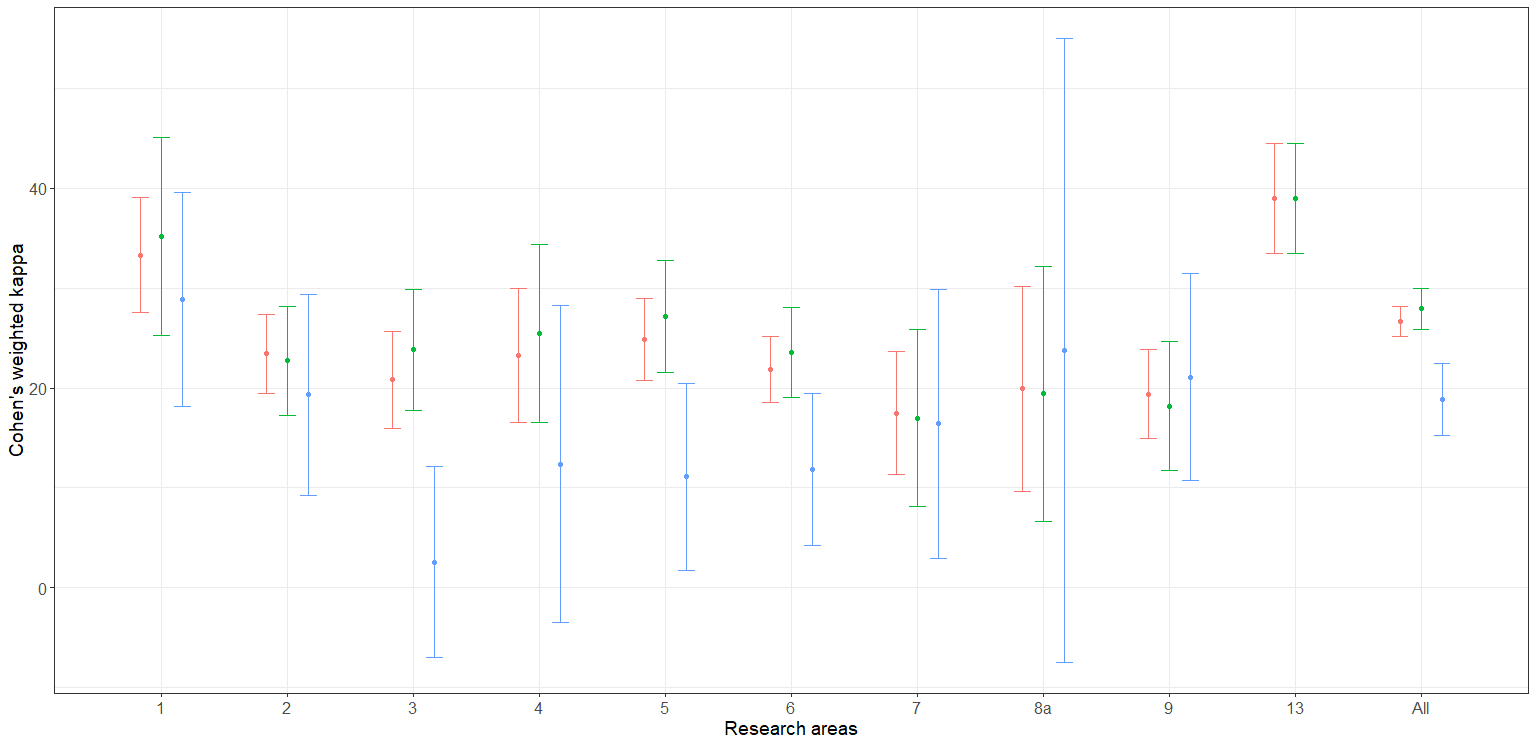} 
\caption{“Error-bar'' plots of the Cohen's kappa coefficient estimates (percent) for EXP1, P1 vs P2 ratings. The confidence intervals are at $95\%$ confidence level and estimates corresponding to $\widehat{\kappa}_w^{(1)}$, $\widehat{\kappa}_w^{(1)}$ (DBR) and $\widehat{\kappa}_w^{(1)}$ (IR) are in red, green and blue, respectively.}
\label{fig:IE_ECON}
\end{figure}

\begin{figure}
\centering	
\includegraphics [scale=0.30]{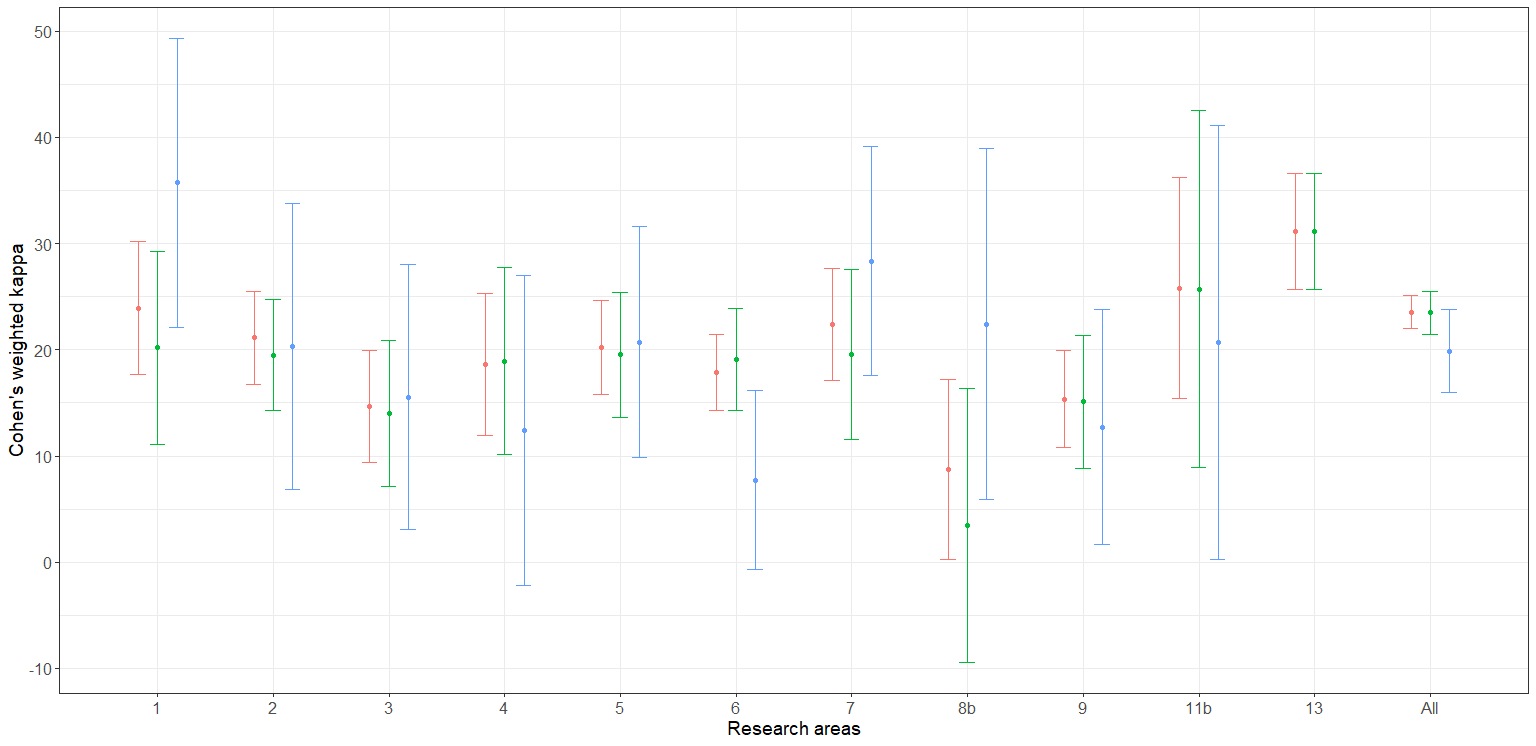} 
\caption{“Error-bar'' plots of the Cohen's kappa coefficient estimates (percent) for EXP2, P1 vs P2 ratings. The confidence intervals are at $95\%$ confidence level and estimates corresponding to $\widehat{\kappa}_w^{(1)}$, $\widehat{\kappa}_w^{(1)}$ (DBR) and $\widehat{\kappa}_w^{(1)}$ (IR) are in green, blue and red, respectively.}
\label{fig:IE_ECON}
\end{figure}

For both experiments, ANVUR directly compared the concordances between P1 and P2 with the ones between peer review and bibliometrics. As for EXP1, Ancaiani et al. (2015) commented the results of the comparison as follows: “the degree of concordance among different reviewers is generally lower than that obtained between the aggregate peer review and the bibliometric evaluation: in this sense, combining evaluations obtained with peer review and bibliometric methods can be considered as more reliable than the usual practice of combining two or more different evaluations obtained by various reviewers of the same article".
Actually, they compared the level of agreement between bibliometrics and peer review (i.e. column ANVUR in Table 3) with the agreement of the two referees for the sub-population DBR (more precisely, column ANVUR in Table 5). When the appropriate estimates are considered, i.e. the second column in Table 5, it is apparent that Ancaiani et al. (2015) statement is no longer true. Hence, their policy suggestion cannot be considered as evidence-based. Actually Ancaiani et al statement appears true only for Area 13, where the concordance indexes between bibliometrics and peer review are much higher than the corresponding indexes between the two reviewers (see Tables 3 and 5). Also in this case, the exception of Area 13 is due to the modification of the protocol of the experiment that boosted the agreement between peer review and bibliometrics.
 
As to EXP2, the agreement between the two reviewers is similar to the agreement between bibliometrics and peer review -- even in Area 13 where the experiment was implemented with a protocol identical to the other areas. These estimates are at odds with ANVUR (2017) conclusions: “It is particularly important the result that the degree of agreement between the bibliometric and the peer evaluation is always higher than the one existing between the two individual peer reviews". Also in this case, ANVUR conclusions were based on estimates computed on the sub-population of articles that boosted -- as previously remarked -- the values of agreement between bibliometrics and peer review.

\section{Testing homogeneity of missing proportions between strata}

In the case of EXP2, Section 4 considers the sizes of missing peer ratings as fixed and -- accordingly --  a design-based approach for the estimation of rating agreement is carried out. However, it could be also interesting to assess the
homogeneity of missing proportions in the different areas by assuming a
random model for the missing peer ratings, i.e. by considering a model-based
approach for missing proportion estimation and testing. In order to provide an appropriate setting in such a case, let us
suppose in turn a fixed population of $N$ items partitioned into $L$ strata.
Moreover, a stratified sampling design is adopted and the notations
introduced in Section 2 are assumed. Hence, each item in the
$h$-th stratum may be missed with probability $\theta_h\in[0,1]$ --
independently with respect to the other items. Thus, the size of missing
items in the $h$-th stratum, say $M_h$, is a random variable (r.v.)
distributed according to the Binomial law with parameters $N_h$ and
$\theta_h$, i.e. the probability function (p.f.) of $M_h$ turns out to be
\begin{equation*}
p_{M_h}(m)=\binom{N_h}{m}\theta_h^m(1-\theta_h)^{N_h-m}\boldsymbol{1}_{\{0,1,%
\ldots,N_h\}}(m)\text{ .}
\end{equation*}

Let us assume that the r.v. $X_h$ represents the size of missing items of the
$h$-th stratum in the sample. By supposing that the items are missing
independently with respect to the sampling design, the distribution of the
r.v. $X_h$ given the event $\{M_h=m\}$ is the Hypergeometric law with
parameters $n_h$, $m$ and $N_h$, i.e. the corresponding conditioned p.f. is
given by
\begin{equation*}
p_{X_h|\{M_h=m\}}(x)=\frac{\binom{m}{x}\binom{N_h-m}{n_h-x}}{%
\binom{N_h}{n_h}}\,\boldsymbol{1}_{\{\max(0,n_h-N_h+m),\ldots,\min(n_h,m)%
\}}(x)\text{ .}
\end{equation*}
Hence, on the basis of this finding and by using the result by Johnson et al.
(2005, p.377), the r.v. $X_h$ is distributed according to the Binomial law
with parameters $n_h$ and $\theta_h$, i.e. the p.f. of $X_h$ is
\begin{equation*}
p_{X_h}(x)=\binom{n_h}{x}\theta_h^x(1-\theta_h)^{n_h-x}\boldsymbol{1}_{\{0,1,%
\ldots,n_h\}}(x)
\end{equation*}
for each $h=1,\ldots,L$. Obviously, the $X_h$'s are independent r.v.'s.

Under the frequentist paradigm, let us consider the null hypothesis of missing proportion homogeneity
$H_0:\theta_h=\theta,$ $\forall h=1,\ldots,L,$ versus the alternative
hypothesis $H_1:\theta_h\neq\theta,\exists h=1,\ldots,L$. For a given
$(x_1,\ldots,x_L)\in\mathbb{N}^L$ such that $y=\sum_{h=1}^Lx_h$, the
likelihood function under the null hypothesis is given by
\begin{equation*}
L_0(\theta)\propto\theta^y(1-\theta)^{n-y}\boldsymbol{1}_{[0,1]}(\theta)%
\text{ ,}
\end{equation*}
while the likelihood function under the alternative hypothesis is$\,$given by
\begin{equation*}
L_1(\theta_1,\ldots,\theta_L)\propto\prod_{h=1}^L\theta_h^{x_h}(1-%
\theta_h)^{n_h-x_h}\boldsymbol{1}_{[0,1]^L}(\theta_1,\ldots,\theta_L)\text{ .}
\end{equation*}
Thus, the likelihood estimator of $\theta$ under the null hypothesis turns
out to be $\widehat{\theta}=Y/n$, where $Y=\sum_{h=1}^LX_h$. In addition, the
likelihood estimator of $(\theta_1,\ldots,\theta_L)$ under the alternative
hypothesis turns out to be $(\widehat{\theta}_1,\ldots,\widehat{\theta}_L)$,
where $\widehat{\theta}_h=X_h/n_h$. 

The likelihood-ratio test statistic could be adopted in order to assess the
null hypothesis. However, in the present setting the large-sample results are
precluded, since the sample size $n$ is necessarily bounded by $N$ and the
data sparsity could reduce the effectiveness of the large-sample
approximations. A more productive approach may be based on conditional
testing (see e.g. Lehmann and Romano, 2005, Chapter 10). First, it is considered the $\chi^2$ test statistic -- asymptotically
equivalent in distribution to the likelihood-ratio test statistic -- which in
this case, after some algebra, reduces to
\begin{equation*}
R:=R(X_1,\ldots,X_L)=\sum_{h=1}^L\frac{n_h(\widehat{\theta}_h-\widehat{%
\theta})^2}{\widehat{\theta}(1-\widehat{\theta})}\text{ .}
\end{equation*}
It should be remarked that the r.v. $Y$ is sufficient for $\theta$ under the
null hypothesis. Hence, the distribution of the random vector
$(X_1,\ldots,X_L)$ given the event $\{Y=y\}$ does not depend on $\theta$.
Moreover, under the null hypothesis, the distribution of the random vector
$(X_1,\ldots,X_L)$ given the event $\{Y=y\}$ is the multivariate
hypergeometric law with parameters $y$ and $(n_1,\ldots,n_L)$, i.e. the
corresponding conditioned p.f. is
\begin{equation*}
p_{(X_1,\ldots,X_L)|\{Y=y\}}(x_1,\ldots,x_L)=\frac{\prod_{h=1}^L%
\binom{n_h}{x_h}}{\binom{n}{y}}\,\boldsymbol{1}_A(x_1,\ldots,x_L)\text{ ,}
\end{equation*}
where
$A=\{(x_1,\ldots,x_L)\in\mathbb{N}^L:x_h\in\{\max(0,n_h-n+y),\ldots,\min(n_h,%
y)\},$ $\sum_{h=1}^Lx_h=y\}$. Thus, by assuming the conditional approach,
an exact test may be carried out. Indeed, if $r$ represents the observed
realization of the test statistic $R$, the corresponding $P$-value is
\begin{equation*}
P(R\geq r\mid\{Y=y\})=\sum_{(x_1,\ldots,x_L)\in
C_r}p_{(X_1,\ldots,X_L)|\{Y=y\}}(x_1,\ldots,x_L)\text{ ,}
\end{equation*}
where $C_r=\{(x_1,\ldots,x_L)\in A:R(x_1,\ldots,x_L)\geq r\}$. It should be
remarked that the previous $P$-value may be approximated by means of a Monte
Carlo method by generating realizations of a Hypergeometric random vector
with parameters $y$ and $(n_1,\ldots,n_L)$. The generation of each
realization requires $(L-1)$ Hypergeometric random variates -- for which
suitable algorithms exist -- and hence the method is practically feasible.

Alternatively, under the Bayesian paradigm, the missing probability homogeneity between
strata may be specified as the model $\mathcal{M}_0$ which assumes that $X_l$
is distributed according to the Binomial law with parameters $n_l$ and
$\theta$, for $l=1,\ldots,L$. In contrast, the model $\mathcal{M}_1$ under
the general alternative postulates that $X_l$ be distributed according to the
Binomial law with parameters $n_l$ and $\theta_l$, for $l=1,\ldots,L$. By
assuming prior distributions in such a way that $\theta$ is elicited as the
absolutely-continuous r.v. $\varTheta$ defined on $[0,1]$ with probability
density function (p.d.f.) given by $f_\varTheta$, while
$(\theta_1,\ldots,\theta_L)$ is elicited as the vector
$(\varTheta_1,\ldots,\varTheta_L)$ of absolutely-continuous r.v.'s defined on
$[0,1]^L$ with joint p.d.f. given by $f_{\varTheta_1,\ldots,\varTheta_L}$,
the Bayes factor may be given by
\begin{equation*}
\begin{aligned}[t]
B_{1,0}&=\frac{\int_{[0,1]^L}\left\{\prod_{_{l=1}}^L\binom{n_l}{x_l}%
\theta_l^{x_l}(1-\theta_l)^{n_l-x_l}\right\}f_{\varTheta_1,\ldots,%
\varTheta_L}(\theta_1,\ldots,\theta_L)d\theta_1\ldots
d\theta_L}{\int_{[0,1]}\left\{\prod_{_{l=1}}^L\binom{n_l}{x_l}\theta^{x_l}(1-%
\theta)^{n_l-x_l}\right\}f_\varTheta(\theta)d\theta}\\
&=\frac{\int_{[0,1]^L}\prod_{_{l=1}}^L\theta_l^{x_l}(1-\theta_l)^{n_l-x_l}f_{%
\varTheta_1,\ldots,\varTheta_L}(\theta_1,\ldots,\theta_L)d\theta_1\ldots
d\theta_L}{\int_{[0,1]}\theta^y(1-\theta)^{n-y}f_\varTheta(\theta)d\theta}%
\text{ .}
\end{aligned}
\end{equation*}
If conjugate priors are considered, the r.v. $\varTheta$ is assumed
distributed according to the Beta law with parameters $a$ and $b$, while
$(\varTheta_1,\ldots,\varTheta_L)$ is the vector of r.v.'s with independent
components, in such a way that each $\varTheta_l$ is distributed according to
the Beta law with parameters $a_l$ and $b_l$. It is worth noting that -- in a
similar setting -- a slightly general hierarchical model is considered by Kass
and Raftery (1995) (see also Albert, 2009, p.190). Hence, the Bayes factor
reduces to
\begin{equation*}
B_{1,0}=\frac{B(a,b)}{B(y+a,n-y+b)}\,\prod_{l=1}^L\frac{B(x_l+a_l,%
n_l-x_l+b_l)}{B(a_l,b_l)}\text{ ,}
\end{equation*}
where -- as usual -- $B(a,b)$ denotes the Euler's Beta function with parameters
$a$ and $b$. In the case of non-informative Uniform priors, i.e. when $a=b=1$
and $a_l=b_l=1$ for $l=1,\ldots,L$, it is apparent that $B_{1,0}$ simplifies
to
\begin{equation*}
B_{1,0}=\frac{\prod_{_{l=1}}^LB(x_l+1,n_l-x_l+1)}{B(y+1,n-y+1)}\text{ .}
\end{equation*}

The testing procedures developed above is applied to the data of EXP2 by considering the areas as the strata 
(see Table 2). At first, by
assuming the frequentist paradigm, the null hypothesis
$H_0$ of missing proportion homogeneity between strata is considered. The null hypothesis
$H_0$ can be rejected since the $P$-value corresponding to the test statistic
$R$ was less than $10^{-16}$. Subsequently, by assuming the Bayesian paradigm
and non-informative Uniform priors, the Bayes factor is computed. In
turn, the missing proportion homogeneity is not likely, since $B_{1,0}$ was
less than $10^{-16}$.
Thus, the conclusions are as follows. Actually, the adoption of stratified random sampling in EXP2 was a suitable design choice, since the population of articles has a structural partition into areas. However, missing data occurred in the stratified sample, since some reviewers refused to referee the assigned articles. Even if this issue is disturbing, it would be a minor drawback if the items were proportionally missed with respect to the strata. Indeed, in such a case it could be argued that the phenomenon is intrinsic in EXP2 -- owing to the different implementation of EXP2 with respect to EXP1. Unfortunately, on the basis of the previous results, the articles are not proportionally missed between the areas, but they are missed according to an unknown random mechanism. As a consequence, the estimates for EXP2 should be considered very carefully, since in some areas (e.g. Areas 6 and 9) the estimated proportion of missing articles is much more elevate with respect to the other areas. Actually, the reasons for which reviewers refused to handle the articles -- or to provide the score in the required time -- are not known and this issue could introduce a further bias in the results of the assessment.

\section{Discussion and conclusion}

The Italian governmental agency for research evaluation ANVUR conducted two experiments for assessing the degree of agreement between bibliometrics and peer review. They were based on stratified random samples of articles, which were classified by bibliometrics and by informed peer review. Subsequently, concordance measures were computed between the ratings resulting from the two evaluation techniques. 
The aim of the two experiments was “to validate the dual system of evaluation" (Ancaiani et al., 2015) adopted in the research assessments. Indeed, in a nutshell, ANVUR used preferentially bibliometric indicators for evaluating articles in the research assessment exercises. When bibliometric rating was inconclusive, ANVUR commissioned a pair of reviewers to evaluate an article: indeed for these articles peer-review evaluation substituted bibliometrics. Bibliometric and peer reviewer ratings were then summed up for computing the aggregate score of research fields, departments and institutions. 
The “dual system of evaluation" might have introduced major biases in the results of the research assessments if bibliometrics and peer review generated systematically different scores. 
A high level of agreement is a necessary condition for  the robustness of research assessment results. The two experiments were designed to test the degree of agreement between bibliometrics and peer review at an individual article level.

%ANVUR computed the results of both experiments after having dropped from the stratified random sample the groups of articles for which the bibliometric algorithms did not produce a score. In addition, as to EXP2, ANVUR also dropped a group of articles for which peer review was unavailable. Therefore, ANVUR presented the results of EXP1 and EXP2 without considering the inferential problems arising from the sample reductions and the missing data. 

This paper reconsiders in full the raw data of the two experiments by adopting the same concordance measure -- i.e. the weighted Cohen's kappa coefficient -- and also the same systems of weights used in EXP1 and EXP2. 
In view of analyzing the experiments in the appropriate inferential setting, the design-based estimation of the Cohen's kappa coefficient and the corresponding confidence interval were developed and adopted for computing the agreement between bibliometrics and peer review in EXP1 and EXP2. Three suggestions are proposed for defining in a proper way the population Cohen's kappa coefficients to be estimated. In a case, the suggested definition represents the suitable version of the coefficient estimated by ANVUR. The other two definitions are advisable for taking into account the sizes of discarded articles by ANVUR.

As to the agreement between bibliometrics and peer review  in EXP1, the point and interval estimates of the considered versions of the weighted Cohen's kappa indicate a concordance degree that can be considered -- at most -- weak, for the aggregate population and for each scientific area. In EXP2 the degree of agreement between bibliometrics and peer review is generally even lower than in EXP1.

Results for Area 13, i.e. Economics and Statistics, deserve a separate consideration. In EXP1, Cohen's  kappa coefficient was estimated to be $54.17\%$. According to Baccini and De Nicolao (2016a), this anomalous high value was possibly due to the modification of the experiment protocol in this area. Indeed, in EXP2 -- when an identical protocol was adopted for all the areas -- the agreement for Area 13 was slightly larger, even if comparable with the other areas. 

Two further points have to be considered. First, the registered lower agreement in EXP2 was arguably due to the adopted systems of ratings, which are based on four categories in EXP1 and on five categories in EXP2. Second, the systems of weights developed by ANVUR tended to boost the value of the weighted Cohen's kappa coefficients with respect to other, more usual, systems of weights (see the supplementary materials providing the computations for linear weights). Hence, the estimates indicate that the “real” level of concordance between bibliometrics and peer review is likely to be worse than weak in both EXP1 and EXP2.

The two experiments also investigated the agreement between the two reviewers, when they score each article of the stratified random sample. 
%Also in this case, in both experiments, ANVUR computed the results after having dropped the articles without a definite bibliometric score. In EXP2, ANVUR dropped also articles for which peer review was unavailable. 
For EXP1, the correct version of the estimates for the article population indicates that the agreement between the two reviewers tend to be lower than $0.30$. A slightly lower concordance level is even obtained for EXP2. In sum, the agreement between pairs of reviewers is weak. In turn, Area 13 represented an exception with the highest level of agreement in both experiments. As previously remarked, in contrast with the other areas, Area 13 adopted a ranking of journals for bibliometric evaluation. When peer reviewers were asked to evaluate a paper, they knew the ranking of journals. Thus, it is possible to conjecture that this very simple information boosted the agreement between reviewers, since they tended to adopt the ranking of journals as a criterion for evaluating articles.

In sum, the two Italian experiments gives concordant evidence that bibliometrics and peer review have less than weak level of agreement at an individual article level. This result is actually consistent with the \textit{Metric Tide} results (Wilsdon et al., 2015). Furthermore, they also show that the agreement between two peer reviewers is in turn very weak. If the agreement between reviewers is interpreted as an estimate of “peer review uncertainty" (Traag and Walman, 2019), this uncertainty is of the same order of magnitude of the uncertainty generated by the use of bibliometrics and peer review.

As to EXP2, a further problem arose for the presence of missing values originated by the refusal of some peer reviewers to referee articles of the sample. For EXP2, the results cannot be easily extended even to the population of journal articles submitted to the research assessment. 

From the evidence presented in this paper, it is possible to carry out a couple of research policy considerations. The first deals with the Italian research assessments exercises. Results of the experiments cannot be considered at all as validating the use of the dual method of evaluation adopted by ANVUR. At the current state of knowledge, it cannot be excluded that the use of the dual method introduced uncontrollable major biases in the final results of the assessments. Indeed, bibliometrics and peer review show a weak agreement. In particular, the evidence drawn from data in the official research reports (ANVUR, 2013, 2017) shows that peer reviewers' scores were on average lower than bibliometric ones. Unbiased results at an aggregate level would be produced solely if the distribution of articles evaluated by the two methods was homogenous for the various units of assessment (research field, research area, departments and universities). Official reports show that the distribution was not homogenous.
\footnote{The distributions per research areas of the articles with an inconclusive bibliometric score and consequently evaluated by peer review varied from 0.9\% to 26.5\% in VQR1 (source: ANVUR, 2013, Table 3.5), and from 0.1\% to 19.2\% in VQR2 (source: ANVUR, 2017, Table 3.5).} Therefore, the aggregate results for research fields, departments and universities might be affected by the proportion of research outputs evaluated by the two different techniques: the higher the proportion of research outputs evaluated by peer review, the lower the aggregate score. From publicly available data, it is possible to show that the average score at the research area level has -- rather generally -- a negative association with the percentage of papers evaluated by peer review. This issue actually holds for VQR1 and VQR2, as shown in the Supplementary materials available at https://doi.org/10.5281/zenodo.3727460. These considerations do not permit to exclude that the results of two Italian research assessments are biased. As a consequence, it is questionable their use for policy purposes and funding distribution. 

Generally, the lesson from the two Italian experiments is that the use of a dual method of evaluation in the same research assessment exercise should be at least considered with extreme caution. A low agreement between bibliometrics and peer review at the level of individual article indicates that metrics should not replace peer review at the level of individual article. The use of the dual methods for reducing costs of evaluation, might dramatically worsen the quality of information obtained in a research assessment exercise.

\newpage

\section*{References}
\vspace{3 mm} Abramo, G. and D’Angelo, C.A. (2016). Refrain from adopting the combination of citation and journal metrics to grade publications, as used in the Italian national research assessment exercise (VQR 2011–2014). \textit{Scientometrics}, 109(3), 2053-2065.
\vspace{3 mm} \\Abramo, G. and D’Angelo, C.A. (2017). On tit for tat: Franceschini and Maisano versus ANVUR regarding the Italian research assessment exercise VQR 2011–2014. \textit{Journal of Informetrics}, 11(3), 783-787.
\vspace{3 mm} \\ Albert, J. (2009). \textit{Bayesian Computation with R} (2nd ed.). New York: Springer.
\vspace{3 mm} \\Alfò, M., Benedetto, S., Malgarini, M. and Scipione, S. (2017). On the use of bibliometric information for assessing articles quality: an analysis based on the third Italian research evaluation exercise. Presentation at the STI 2017, September 6-8, 2017, Paris.
\vspace{3 mm} \\Ancaiani, A., Anfossi, A.F., Barbara, A., Benedetto, S., Blasi, B., Carletti, V., Cicero, T., Ciolfi, A., Costa, F., Colizza, G., Costantini, M., di Cristina, F., Ferrara, A., Lacatena, R.M., Malgarini, M., Mazzotta, I., Nappi, C.A., Romagnosi, S. and Sileoni, S. (2015). Evaluating scientific research in Italy: The 2004–10 research evaluation exercise. \textit{Research Evaluation}, 24(3), 242-255.
\vspace{3 mm} \\ANVUR (2013). \textit{Valutazione della qualità della ricerca 2004-2010. Rapporto finale}.
\vspace{3 mm} \\ANVUR (2017). \textit{Valutazione della qualità della ricerca 2011-2014. Rapporto finale}.
\vspace{3 mm} \\Baccini, A. and De Nicolao, G. (2016a). Do they agree? Bibliometric evaluation versus informed peer review in the Italian research assessment exercise. \textit{Scientometrics}, 108(3), 1651-1671.
\vspace{3 mm} \\Baccini, A. and De Nicolao, G. (2016b). Reply to the comment of Bertocchi et al.  \textit{Scientometrics}, 108(3), 1675-1684.
\vspace{3 mm} \\Baccini, A. and De Nicolao, G. (2017a). A letter on Ancaiani et al. ‘Evaluating scientific research in Italy: the 2004-10 research evaluation exercise’. \textit{Research Evaluation}, 26(4), 353-357.
\vspace{3 mm} \\Baccini, A. and De Nicolao, G. (2017b). Errors and secret data in the Italian research assessment exercise. A comment to a reply. \textit{RT. A Journal on Research Policy and Evaluation}, 5(1).
\vspace{3 mm} \\Benedetto, S., Cicero, T., Malgarini, M. and Nappi, C.S. (2017). Reply to the letter on Ancaiani et al. ‘Evaluating scientific research in Italy: The 2004–10 research evaluation exercise’. \textit{Research Evaluation}, 26(4), 358-360.
\vspace{3 mm} \\Berry, K.J., Johnston, J.E. and Mielke, P.W. (2018). \textit{The Measurement of
Association}. Switzerland: Springer Nature.
\vspace{3 mm} \\Bertocchi, G., Gambardella, A., Jappelli, T., Nappi, C.A. and Peracchi, F. (2013). Bibliometric evaluation vs. informed peer review: Evidence from Italy. Unpublished manuscript. Naples: CSEF Working Papers.
\vspace{3 mm}\\Bertocchi, G., Gambardella, A., Jappelli, T., Nappi, C.A. and Peracchi, F. (2015). Bibliometric evaluation vs. informed peer review: Evidence from Italy. \textit{Research Policy}, 44(2), 451-466.
\vspace{3 mm}\\Bertocchi, G., Gambardella, A., Jappelli, T., Nappi, C.A. and Peracchi, F. (2016). Comment to: do they agree? Bibliometric evaluation versus informed peer review in the Italian research assessment exercise. \textit{Scientometrics}, 108(1), 349-353.
\vspace{3 mm} \\Cohen, J. (1960). A coefficient of agreement for nominal scales.  \textit{Educational and Psychological Measurement}, 20(1), 37-46.
\vspace{3 mm} \\Cohen, J. (1968). Weighted kappa: nominal scale agreement with provision for scaled disagreement or partial credit. \textit{Psychological Bulletin}, 70(4), 213-220.
\vspace{3 mm} \\De Raadt, A., Warrens M.J., Bosker, R.J. and Kiers, H.A.L. (2019). Kappa
coefficients for missing data. \textit{Educational and Psychological Measurement}, 79(3),
558-576.
\vspace{3 mm} \\Demnati, A. and Rao, J.N.K. (2004). Linearization variance estimators for survey data (with discussion). \textit{Survey Methodology}, 30(1), 17-34.
\vspace{3 mm} \\Fagerland, M.W., Lydersen, S. and Laake, P. (2017). \textit{Statistical Analysis of Contingency Tables}. Boca Raton: CRC Press.
\vspace{3 mm} \\Fleiss, J.L., Levin, B. and Paik, M.C. (2003). \textit{Statistical Methods for Rates and Proportions} (3rd ed.). Hoboken: Wiley. 
\vspace{3 mm} \\Franceschini, F. and Maisano, D. (2017). Critical remarks on the Italian research assessment exercise VQR 2011–2014. \textit{Journal of Informetrics}, 11(2), 337-357.
%\vspace{3 mm} \\Goukrager, S. (2012).  \textit{Objectivity}. Oxford: Oxford University Press.
\vspace{3 mm} \\Gwet, K.L. (2014). \textit{Handbook of Inter-rater Reliability: the Definitive Guide to Measuring the Extent of Agreement among Multiple Raters}. Gaithersburg: Advanced Analytics.
\vspace{3 mm} \\Hicks, D. (2012). Performance-based university research funding systems. \textit{Research Policy}, 41(2), 251-261.
\vspace{3mm} \\Johnson, N., Kemp, A. and Kotz, S. (2005).  \textit{Univariate Discrete Distributions} (3rd ed.). New York: Wiley.
\vspace{3mm} \\ HEFCE (2015). \textit{The Metric Tide: Correlation analysis of REF2014 scores and metrics (Supplementary Report II to the Independent Review of the Role of Metrics in Research
Assessment and Management)}. HEFCE, DOI: 10.13140/RG.2.1.3362.4162. 
\vspace{3mm} \\Kass, R.E. and Raftery, A.E. (1995). Bayes factor and model uncertainty. \textit{Journal of the American Statistical Association}, 90, 773-795.
\vspace{3 mm} \\Kulczycki, E., Korzeń, M. and Korytkowski, P. (2017). Toward an excellence-based research funding system: Evidence from Poland. \textit{Journal of Informetrics}, 11(1), 282-298.
\vspace{3 mm} \\Lehmann, E.L. and Romano J.P. (2005).  \textit{Testing Statistical Hypotheses} (3rd ed.). New York: Springer.
\vspace{3 mm} \\Pride, D. and Knoth, P. (2018). Peer review and citation data in predicting university rankings, a large-scale analysis.  in: Méndez E., Crestani F., Ribeiro C., David G., Lopes J. (eds.) \textit{Digital Libraries for Open Knowledge}. TPDL 2018. \textit{Lecture Notes in Computer Science}, 11057. Cham: Springer.
\vspace{3 mm} \\Quatember, A. (2015). \textit{Pseudo-Populations: A Basic Concept in Statistical Surveys}. New York: Springer.
\vspace{3 mm} \\R Core Team (2019). \textit{R: A Language and Environment for Statistical Computing}. R Development Core Team, R Foundation for Statistical Computing, Vienna, Austria.
\vspace{3 mm} \\Sheskin, D.J. (2003).  \textit{Handbook of Parametric and Nonparametric Statistical Procedures} (3rd ed.). London: Chapman \& Hall.
\vspace{3 mm} \\Strijbos, J.W., Martens, R.L. Prins, F.J. and Jochems, W.M.G. (2006). Content analysis: What are they talking about? \textit{Computers \& Education}, 46(1), 29-48.
\vspace{3 mm} \\Strijbos, J.W. and Stahl, G. (2007). Methodological issues in developing a multi-dimensional coding procedure for small-group chat communication. \textit{Learning and Instruction}, 17(4), 394-404.
\vspace{3 mm} \\Thompson, M.E. (1997). \textit{Theory of Sample Surveys}. London: Chapman \& Hall.
\vspace{3 mm} \\Traag, V.A. and Waltman, L. (2019). Systematic analysis of agreement between metrics and peer review in the UK REF. \textit{Palgrave Communications}, 5(1), 29.
\vspace{3 mm} \\Uebersax, J.S. (1987). Diversity of decision-making models and the measurement
of interrater agreement. \textit{Psychological Bulletin}, 101(1), 140-146. 
\vspace{3 mm} \\Wilsdon, J., Allen, L., Belfiore, E., Campbell, P., Curry, S., Hill, S., Jones, R., Kain, R., Kerridge, S., Thelwall, M., Tinkler, J., Viney, I., Wouters, P., Hill, J. and Johnson, B. (2015).  \textit{The Metric Tide: Report of the Independent Review of the Role of Metrics in Research Assessment and Management}. HEFCE, DOI: 10.13140/RG.2.1.4929.1363
\vspace{3 mm} \\Wolfram Research, Inc. (2014) \textit{Mathematica}, Version 10.0, Champaign, Illinois.
\vspace{3 mm} \\Wouters, P., Thelwall, M., Kousha, K., Waltman, L., de Rijcke, S., Rushforth, A. and Franssen, T. (2015). \textit{The Metric Tide: Literature Review (Supplementary Report I to the
Independent Review of the Role of Metrics in Research Assessment and Management)}. HEFCE, DOI: 10.13140/RG.2.1.5066.3520.

\end{document}